\documentclass[prx,twocolumn,superscriptaddress]{revtex4-2}

\usepackage{tcolorbox}
\usepackage{graphicx}
\usepackage{mathtools}
\usepackage{amsmath}
\usepackage{amssymb}
\usepackage{xcolor}

\usepackage{braket}

\usepackage[pdfencoding=auto, psdextra]{hyperref}
\hypersetup{
	colorlinks=true, 
	linktoc=all,     
	linkcolor=blue,  
}

\newcommand{\Tr}{\text{Tr}}
\newcommand{\mb}{\mathbf}
\newcommand{\rme}{\mathrm{e}}
\newcommand{\kb}[2]{\ket{#1}\bra{#2}}
\newcommand{\intf}[3]{\int_{#1}^{#2}\text{d}#3\ }
\newcommand{\uk}{_\mb{k}}
\newcommand{\uq}{_\mb{q}}

\newcommand{\proj}[2]{\left|#1\right>\left<#2\right|}

\begin{document}
	
	
	
	
	\title{Optical polaron formation in quantum systems with permanent dipoles} 
	
	\author{Adam Burgess}
	\email[]{a.d.burgess@surrey.ac.uk}
	\affiliation{Leverhulme Quantum Biology Doctoral Training Centre, University of Surrey, Guildford, GU2 7XH, United Kingdom}
	\affiliation{Department of Physics and Advanced Technology Institute,  University of Surrey, Guildford, GU2 7XH, United Kingdom}
	
	\author{Marian Florescu}
	\email[]{m.florescu@surrey.ac.uk}
	\affiliation{Department of Physics and Advanced Technology Institute,  University of Surrey, Guildford, GU2 7XH, United Kingdom}

	\author{ Dominic M. Rouse}
	\email[]{dominic.rouse@glasgow.ac.uk}
	\affiliation{School of Physics and Astronomy, University of Glasgow, Glasgow G12 8QQ, Scotland, United Kingdom}
	\affiliation{Department of Physics and Astronomy, University of Manchester, Oxford Road, Manchester M13 9PL, United Kingdom}
	
	\date{\today}
	\begin{abstract}
		Many optically active systems possess spatially asymmetric electron orbitals. These generate permanent dipole moments, which can be stronger than the corresponding transition dipole moments, significantly affecting the system dynamics and creating polarised Fock states of light. We derive a master equation for these systems by employing an optical polaron transformation that captures the photon mode polarisation induced by the permanent dipoles. This provides an intuitive framework to explore their influence on the system dynamics and emission spectrum. We find that permanent dipoles introduce multiple-photon processes and a photon sideband which causes substantial modifications to single-photon transition dipole processes. In the presence of an external drive, permanent dipoles lead to an additional process that we show can be exploited to optimise the decoherence and transition rates. We derive the emission spectrum of the system, highlighting experimentally detectable signatures of optical polarons, and measurements that can identify the parameters in the system Hamiltonian, the magnitude of the differences in the permanent dipoles, and the steady-state populations of the system.
	\end{abstract}
	
	\maketitle
	\section{Introduction}
	In general, the interaction of atomic systems with light through transition dipoles is well understood, and the optical master equation describing exciton creation and annihilation through photon emission and absorption, respectively, is derived in many introductory texts dedicated  to open quantum systems \cite{breuer2002theory,ficek2005quantum,leggett1987dynamics,agarwal1974quantum}. Atomic systems have highly symmetric electron orbitals and so possess negligible permanent dipoles. However, many physical systems do not share this property and can possess permanent dipoles stronger than their transition dipole moments. Such systems include molecules with parity mixing of the molecular state \cite{chung2016determining,filippi2012bathochromic,kovarskiui2001effect,deiglmayr2010permanent,guerout2010ground,lin2020mechanism,jagatap2002contributions,gilmore2005spin}, quantum dots with asymmetric confining potentials \cite{garziano2016one,chestnov2016ensemble,shim1999permanent,anton2016radiation,fry2000photocurrent,chestnov2017terahertz,fry2000inverted,anton2017optical,patane2000piezoelectric,warburton2002giant,warburton2002giant,ostapenko2010large}, nanorods with non-centrosymmetric crystallographic lattices \cite{li2003origin,gupta2006self,mohammadimasoudi2016full}, and superconducting circuits \cite{yoshihara2017superconducting}. 
	
	Permanent dipoles introduce additional pure dephasing interactions into the Hamiltonian. The non-additivity of the pure dephasing and transition dipole interactions yields unique physical effects, including modifications to decoherence \cite{guarnieri2018steady,greenberg2007low}, steady-state coherence \cite{guarnieri2018steady,roman2021enhanced,purkayastha2020tunable}, laser-driven population inversion \cite{macovei2015population}, multiphoton conversion \cite{mirzac2021microwave,mandal2020polarized}, entanglement generation \cite{anton2020bichromatically,oster2012generation}, and second-harmonic generation \cite{juzeliunas2003eliminating,paspalakis2013effects}. Understanding the role of strong permanent dipoles is highly relevant for the design of new quantum technologies, as well as in exploring novel biochemical processes. Previous studies have either neglected the transition dipole moments assuming only a pure dephasing interaction  \cite{gilmore2005spin}, have considered single-mode fields \cite{greenberg2007low,zhao2016effect,hattori1987bloch,scala2021beyond,mandal2020polarized}, or have treated the permanent dipoles in a perturbative manner \cite{gilmore2005spin,guarnieri2018steady,anton2020bichromatically,greenberg2007low,zhao2016effect,hattori1987bloch,scala2021beyond}. Furthermore, most studies neglect an additional identity interaction also induced by the permanent dipoles, which, as we show here, modifies the initial state of the environment and can have a significant impact on the system dynamics.  Consequently, such treatments do not capture the role of strong permanent dipoles in asymmetric systems under illumination by multimode fields, such as common thermal fields.

	
	In this paper, we utilise a polaron transformation to derive a master equation for systems with strong permanent dipoles interacting with a thermal field. Importantly, we make no assumptions about the dipole matrix beyond perturbative \textit{transition} dipoles. The polaron transformation is a unitary, state-dependent displacement transformation widely used when dealing with strong, pure dephasing interactions. 
	In the polaron frame, pure dephasing interactions are absorbed into the definition of the basis and treated to all orders in the coupling strength. The basis describes an optical polaron quasiparticle, which is a hybridisation of the matter excitations and the displaced harmonic oscillator states of the multi-mode photonic field. These photonic states, called polarised Fock states, correspond to a non-zero (polarised) vector potential field and were first introduced to explain how permanent dipoles can generate multiphoton conversion in a single-mode cavity \cite{mandal2020polarized} and are useful in polaritonic chemistry \cite{mandal2020polarized2,mandal2023theory}.
	
	The optical polaron formalism provides us with an essential intuition for the very complex phenomena introduced by the permanent dipoles, allowing us to unpick the role of permanent dipoles in open quantum system dynamics. We find that the formation of optical polarons results in unique physical phenomena such as modifications to single-photon transition dipole processes, entirely new multiple photon processes, the appearance of photonic sidebands, and a novel interplay with external driving that allows control over the dynamics. We also derive the emission spectrum for the system, highlighting experimentally detectable signatures of the optical polarons.


	
	
	This paper is organised as follows. In Sec.~\ref{sec:model}, we introduce the Hamiltonian and transform it into the polaron frame. In Sec.~\ref{sec:ECF}, we derive the polaron frame master equation and discuss the new physical processes using the analytical expressions. Following this, in Sec.~\ref{sec:ME}, we compare the polaron frame master equation to perturbative dynamics and to numerically exact dynamics using the time-evolving matrix product operators (TEMPO) \cite{strathearn2017efficient,Strathearn2018,pollock2018non,jorgensen2019exploiting,gribben2022exact,fux2021efficient,gribben2022using,fux2022thermalization,fowler2022efficient} algorithm, through the open source code OQuPy \cite{oqupy}. In Sec.~\ref{sec:spectra} we derive the emission spectrum, and in Sec.~\ref{sec:IC} we briefly discuss the role of the initial state and identity type interactions. Finally in Sec.~\ref{sec:conc} we present concluding remarks.

	\section{The model and polaron frame}\label{sec:model}
	
	We consider a driven asymmetric emitter with a single quantised dipole coupled to a long wavelength multimode cavity. After truncating the material subsystem to its two lowest energy levels, the fundamental multipolar-gauge Hamiltonian is
	\begin{equation}\label{eq:H}
		H=\frac{\epsilon}{2}\sigma_z+V\sigma_++V^*\sigma_-+\sum_k\nu\uk a_k^\dagger a_k + \mb{d}\cdot\mb{\Pi}+E_\text{dip},
	\end{equation}
	where $\epsilon$ is the transition energy of the quantum emitter and $V=|V|\rme^{i\vartheta_V}$ is a complex drive. $\mb{\Pi}=i\sum_k \mb{e}_k f\uk( a_k^\dagger -a_k )$ is the electric displacement field where  $k=\{\mb{k},\lambda\}$ is a four-vector representing both the wavevector $\mb{k}$ and polarisation state $\lambda$ of the mode with polarisation vector  $\mb{e}_k$, energy $\nu\uk$, and coupling strength $f\uk=\sqrt{\nu\uk/2\mathcal{V}}$ where $\mathcal{V}$ is the field volume. $a_{k}$ and $a^{\dagger}_{k}$ are the field mode annihilation and creation operators \cite{babiker1983derivation,stokes2018master,mahan2013many}. The dipolar self energy term is
	\begin{equation}
		E_\text{dip}=\sum_k\frac{f\uk^2}{\nu\uk}\left(\mb{d}\cdot\mb{e}_k\right)^2.
	\end{equation}
	In the truncated system Hilbert space, the dipole operator is
	\begin{align}\label{eq:d}
		\mb{d}&=\begin{pmatrix}
			\mb{d}_{ee} & \mb{d}_{eg}\\
			\mb{d}_{ge} & \mb{d}_{gg}
		\end{pmatrix}=\mb{d}_\Delta\sigma_z+\mb{d}_D\mathcal{I}+\mb{d}_\mu\sigma_++\mb{d}_\mu^*\sigma_-,
	\end{align}
	where the Pauli operators are $\sigma_z=\kb{e}{e}-\kb{g}{g}$, $\sigma_+=\kb{e}{g}$, $\sigma_-=\kb{g}{e}$, $\mathcal{I}=\kb{g}{g}+\kb{e}{e}$, and $\mb{d}_{ij}=\langle i|\mb{d}|j\rangle$ for $i,j\in\{e,g\}$. We have defined the following combinations of dipole matrix elements,
	\begin{align}
		\mb{d}_\Delta=\frac{\mb{d}_{ee}-\mb{d}_{gg}}{2},\quad
		\mb{d}_D=\frac{\mb{d}_{ee}+\mb{d}_{gg}}{2},\quad
		\mb{d}_\mu=\mb{d}_{eg},
	\end{align}
	which play an essential role in our analysis. The $\mb{d}_{p}$ vectors are not guaranteed to be co-linear and Hermiticity of $H$ requires that $\mb{d}_{p}\in\mathbb{R}^3$ for $p\in\{\Delta,D\}$ and $\mb{d}_\mu\in\mathbb{C}^3$.
	
	At this point it is often assumed that either $|\mb{d}_{\Delta}|\approx|\mb{d}_{D}|\approx0$ which leads to the standard optical master equation \cite{breuer2002theory,ficek2005quantum,leggett1987dynamics,agarwal1974quantum}, or that $|\mb{d}_{D}|\approx|\mb{d}_\mu|\approx0$ leading to a pure dephasing interaction \cite{gilmore2005spin}. In both of those limits, the dipolar self energy term is proportional to the identity. In this study, we make no assumptions about the size of the permanent dipoles.
	
	Substituting Eq.~\eqref{eq:d} into Eq.~\eqref{eq:H} and absorbing the drive phase into a basis $\ket{e'}=\rme^{i\vartheta_V/2}\ket{e}$ and $\ket{g'}=\rme^{-i\vartheta_V}\ket{g}$ we find
	\begin{align}\label{eq:Hl}
		H=&\ \frac{\epsilon}{2}\sigma_z'+|V|\sigma_x'+\sum_k\nu\uk a_k^\dagger a_k+E_\text{dip}+\pi_{DD}\mathcal{I}'\nonumber\\
		&+\pi_{\Delta\Delta}\sigma_z'+\pi_{\mu\bar{\mu}}\rme^{-i\vartheta_V}\sigma_+'+\pi_{\bar{\mu}\mu}\rme^{i\vartheta_V}\sigma_-',
	\end{align}
	where primed operators are in the $\{\ket{e'},\ket{g'}\}$ basis and
	\begin{align}
		\pi_{pq}&=\sum_k \left(p_k a_k^\dagger+q_k^*a_k\right)\label{eq:Apq},\\
		p_k&=if\uk\left(\mb{d}_p\cdot\mb{e}_k\right)\label{eq:param},
	\end{align} 
	with $p,q\in\{\mu,\bar{\mu},\Delta,D\}$ and we denote $\mb{d}_{\bar{\mu}}=\mb{d}_\mu^*$. We refer to Eq.~\eqref{eq:Hl} as the lab frame Hamiltonian, which is equivalent to Eq.~\eqref{eq:H}. 
	
	The photon-only part of Eq.~\eqref{eq:Hl} can be diagonalised by the displacement transformation $H_d=B(D/\nu)HB(-D/\nu)$, where subscript $d$ denotes the displaced frame, and displacement operators are given by 
	\begin{equation}
		B(\alpha)=\rme^{\sum_k\left(\alpha_k a_k^\dagger-\alpha_k^* a_k\right)},
	\end{equation}
	and $B(\alpha)^\dagger=B(-\alpha)$. These act on photon operators by
	\begin{equation}\label{eq:disp}
		B(\pm\alpha)a_kB(\mp\alpha)=a_k\mp\alpha_k,
	\end{equation}
	and we further analyse the displacement operators in Appendix~\ref{app:disp}. Ignoring terms proportional to the identity, the resulting Hamiltonian is
	\begin{align}\label{eq:Hd}
		H_d=&\frac{\epsilon}{2}\sigma_z'+|V|\sigma_x'+\sum_k\nu\uk a_k^\dagger a_k\nonumber\\
		&+\pi_{\Delta\Delta}\sigma_z'+\pi_{\mu\bar{\mu}}\rme^{-i\vartheta_V}\sigma_+'+\pi_{\bar{\mu}\mu}\rme^{i\vartheta_V}\sigma_-'.
	\end{align}
	Notice that the dipolar self energy term has cancelled \textit{exactly} with the terms that result from displacing the light-matter interactions, such that Eq.~\eqref{eq:Hd} is independent of $D_k$.
	
	Throughout this paper, we make the standard assumption that the transition dipole moment $\mb{d}_\mu$ is small enough to permit an accurate second-order perturbative expansion in its magnitude $|\mb{d}_\mu|$. Even for perturbative transition dipole moments, the master equation derived using $H_d$ in Eq.~\eqref{eq:Hd}, referred to as the \textit{displaced frame master equation} (DFME), will become inaccurate if $|\mb{d}_\Delta|$ is large. 
	
\begin{figure}[ht!]\centering
		\includegraphics[width=0.48\textwidth]{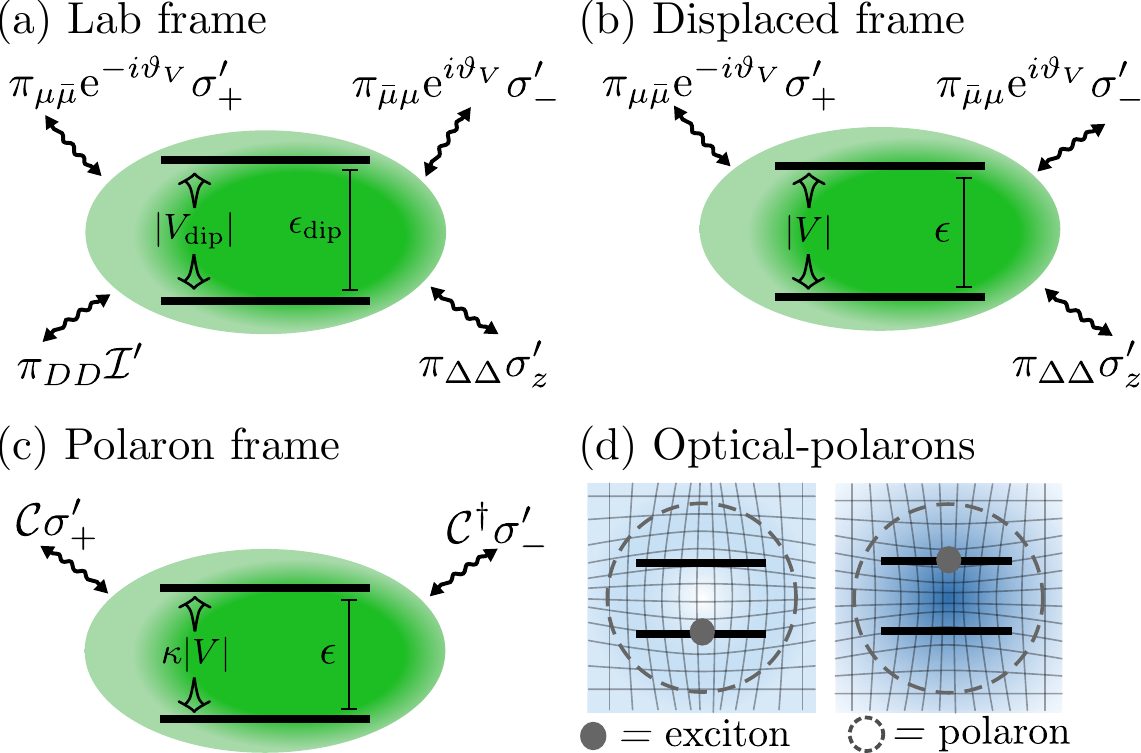}
		\caption{\textbf{Illustration of the model and the optical polaron concept.} (a)-(c) depicts the molecular energies and interactions in the (a) lab frame, (b) displaced frame and (c) polaron frame. In the lab frame, the dipolar self energy term $E_\text{dip}$ causes the renormalisations $\epsilon\to\epsilon_\text{dip}$ and $V\to V_\text{dip}$, which cancel out in the displaced and polaron frames. (d) is an illustration of an optical polaron: a quasiparticle formed of the matter excitation and the polarised Fock states of the displaced photon modes.}
		\label{fig:pol}
	\end{figure}

	To overcome this challenge we make a polaron transformation prior to deriving the Redfield master equation. In this frame, the polarising effect of the pure dephasing interaction on the field is absorbed into the definition of a new basis, called the polarised Fock states \cite{mandal2020polarized,mandal2020polarized2}. The polarisation direction of the Fock state is dependent on the matter state, and it hybridises with the excitation to create an optical polaron quasiparticle. In Fig.~\ref{fig:pol}, we illustrate the various frames introduced and the optical polaron concept. A Redfield master equation derived in the new basis, termed the \textit{polaron frame master equation} (PFME), will be robust to all magnitudes of the permanent dipoles and will recover the DFME in the limit of $\Delta_k\to0$.
	
	The polaron frame Hamiltonian is $H_p=UH_dU^\dagger$,  where
	\begin{equation}\label{eq:U}
		U=B\left(\frac{\Delta}{\nu}\right)\kb{e'}{e'}+B\left(-\frac{\Delta}{\nu}\right)\kb{g'}{g'}.
	\end{equation}
	Using Eq.~\eqref{eq:disp}, and ignoring constant terms, we obtain		\begin{align}\label{eq:Hp}
		H_p=\frac{\epsilon}{2}\sigma_z'+\kappa |V|\sigma_x'+\sum_k\nu\uk a_k^\dagger a_k+\mathcal{C}\sigma_+'+\mathcal{C}^\dagger \sigma_-',
	\end{align}
	where the coupling operator is $\mathcal{C}=C-\langle C\rangle$ with 
	\begin{align}\label{eq:C}
		C=B\left(\frac{\Delta}{\nu}\right)\pi_{\mu\bar{\mu}}B\left(\frac{\Delta}{\nu}\right)\rme^{-i\vartheta_V}+|V|B\left(2\frac{\Delta}{\nu}\right).
	\end{align}
	We denote $\langle \cdot \rangle=\text{Tr}_E(\cdot \rho_E)$ where \begin{equation}\label{eq:rhoB}
		\rho_E=\frac{1}{\mathcal{Z}_E}\rme^{-\beta\sum_k\nu\uk a_k^\dagger a_k},
	\end{equation}
	and $\mathcal{Z}_E=\Tr[\exp(-\beta\sum_k\nu\uk a_k^\dagger a_k)]$ with $\beta=1/(k_B T)$ the inverse temperature. In Appendix~\ref{app:Cexp} we prove that $\langle C \rangle=\kappa |V|$ where $\kappa=\langle B(\pm 2\Delta/\nu)\rangle$. In Eq.~\eqref{eq:Hp} we moved a factor of $\langle C \rangle \sigma_x'$ from the coupling operator and into the unperturbed part of $H_p$. This is to ensure that the perturbation theory yields the Redfield equation.
	
	In the polaron frame Hamiltonian in Eqs.~\eqref{eq:Hp} we can now interpret $\sigma_{\pm}'$ as causing transitions between the optical polaron states in Fig.~\ref{fig:pol}(d). Compared to matter excitations, described by the Hamiltonian in the displaced frame in Eqs.~\eqref{eq:Hd}, polarons experience a more complicated interaction, albeit one without a pure dephasing term. We also note that the polaron frame Hamiltonian is diagonal when the transition dipoles and driving vanish. Hence, the permanent dipoles generate trivial dynamics if considered in isolation, described by the independent boson model.

    Finally, the unperturbed polaron Hamiltonian can be diagonalised using a unitary rotation, $(\epsilon/2)\sigma_z'+ \kappa |V|\sigma_x'=(\eta/2)\tau_z$, where
	\begin{align}
		\eta=&\ \sqrt{\epsilon^{ 2}+ 4\kappa^2\left| V\right|^2},\label{eq:eta}\\
		\tau_z=&\ \kb{+}{+}-\kb{-}{-}.
	\end{align}
	The eigenbasis relates to the original basis by
	\begin{equation}
		\begin{pmatrix}
			\ket{e'} \\ \ket{g'}
		\end{pmatrix}
		=
		\begin{pmatrix}
			\cos\left(\frac{\varphi}{2}\right) & -\sin\left(\frac{\varphi}{2}\right) \\
			\sin\left(\frac{\varphi}{2}\right) & \cos\left(\frac{\varphi}{2}\right)
		\end{pmatrix}
		\begin{pmatrix}
			\ket{+} \\ \ket{-}
		\end{pmatrix},
	\end{equation}
	with $\cos(\varphi)=\epsilon/\eta$ and $\sin(\varphi)=2\kappa|V|/\eta$. In the eigenbasis, the polaron frame Hamiltonian is
	\begin{equation}\label{eq:Ha}
		H_p=\frac{\eta}{2}\tau_z+\sum_k\nu\uk a_k^\dagger a_k+\sum_{\alpha\in\{z,+,-\}}g_\alpha\tau_\alpha,	
	\end{equation}
	where we have defined $\tau_+=\kb{+}{-}$, $\tau_-=\kb{-}{+}$, and the coupling operators
	\begin{subequations}\label{eq:g}
		\begin{align}
			g_z=&\ \frac{1}{2}\sin\left(\varphi\right)\left[\mathcal{C}+\mathcal{C}^\dagger\right],\label{eq:gz}\\
			g_+=&\ \left[\cos^2\left(\frac{\varphi}{2}\right) \mathcal{C}-\sin^2\left(\frac{\varphi}{2}\right) \mathcal{C}^\dagger \right],
		\end{align}
	\end{subequations}
	and $g_-=g_+^{\dagger}$. 
	
	To derive the master equation we will take the continuum limit of the photon modes, in which summations over an arbitrary function $F(\nu)$ transform according to 
	\begin{align}\label{eq:CL1}		\sum_k f\uk^2F(\nu\uk)&\left(\mb{d}_p\cdot\mb{e}_k\right)\left(\mb{d}_q\cdot\mb{e}_k\right)\nonumber\\
		\qquad&\to h_{pq}\intf{0}{\infty}{\nu}J(\nu)F(\nu), 
	\end{align}
	where $p,q\in\{\mu,\bar{\mu},\Delta\}$ and
 \begin{equation}
h_{pq}=\int_{\Omega\uk}\text{d}\Omega\uk\ \sum_\lambda\left(\tilde{\mb{d}}_p\cdot\mb{e}_k\right)\left(\tilde{\mb{d}}_q\cdot\mb{e}_k\right).
 \end{equation} 
 We have defined a spectral density $J(\nu)$, a dimensionless dipole vector $\tilde{\mb{d}}_p=\mb{d}_p/|\mb{d}_\text{ref}|$ measured against an arbitrary reference value $|\mb{d}_\text{ref}|$, and the solid angle integral $\int_{\Omega\uk}\text{d}\Omega\uk=\intf{0}{\pi}{\theta\uk}\sin(\theta\uk)\intf{0}{2\pi}{\phi\uk}$ \cite{ficek2005quantum}. For free space  and   unpolarised light,
 \begin{subequations}\label{eq:gpq}
	\begin{align}\label{eq:Omegaij}	h_{\mu\Delta}&=\Omega_{\mu\Delta}\rme^{i\vartheta_\mu}\cos\left(\theta_{\mu\Delta}\right),\\
 h_{\mu\bar{\mu}}&=\Omega_{\mu\mu},\\
 h_{\mu\mu}&=\Omega_{\mu\mu}\rme^{2i\vartheta_\mu},
	\end{align}
 \end{subequations}
	where $\Omega_{pq}=\Omega_{qp}=(8\pi/3)|\tilde{\mb{d}}_p||\tilde{\mb{d}}_q|\in\mathbb{R}$, $\vartheta_\mu$ is the complex phase of $\mb{d}_\mu$, $\theta_{\mu\Delta}$ is the angle between $\mb{d}_\Delta$ and $\mb{d}_\mu$, and one can obtain e.g. $h_{\bar{\mu}\Delta}=h_{\mu\Delta}^*$ by suitable complex conjugation. Notice that the dynamics only depend on the relative phase $\vartheta_{\mu V}=\vartheta_\mu-\vartheta_V$.
 
 We use the free space multipolar-gauge spectral density
	\begin{equation}\label{eq:SD}
		J(\nu)=S \frac{\nu^3}{\nu_c^2}\rme^{-\frac{\nu}{\nu_c}}\Theta(\nu),
	\end{equation}
	which has dimensions of energy. $\Theta(\nu)$ is the Heaviside step function and $S=\intf{0}{\infty}{\nu}J(\nu)/\nu^2$ is a dimensionless parameter called the Huang-Rhys parameter. We have introduced a phenomenological cut-off to the spectral density $\nu_c$, which is justified for finite systems \cite{stokes2018master}. In the continuum limit, $\kappa=\exp[-\phi(0)/2]$ where the photon propagator is
	\begin{align}\label{eq:phi}
		\phi(s)=4\Omega_{\Delta\Delta}\intf{0}{\infty}{\nu}\frac{J(\nu)}{\nu^2}&\big[\coth\left(\frac{\beta\nu}{2}\right)\cos\left(\nu s\right)\nonumber\\
		&\qquad\qquad-i\sin\left(\nu s\right)\big].
	\end{align}
	
	To summarise, the important parameters in this model are the dipole strengths $|\tilde{\mb{d}}_\mu|$ and $|\tilde{\mb{d}}_\Delta|$, their relative angle $\theta_{\mu\Delta}$, the eigenenergy $\eta$, the driving strength $|V|$, and the relative complex phase $\vartheta_{\mu V}$. We denote by $\varphi$ the angle determining the eigenbasis, by $\theta_{\mu\Delta}$ the angle between the dipole vectors, and by $\vartheta_{\mu V}$ the relative complex phases. Whether or not the transition or permanent dipole interactions are deemed strong is approximately determined by $|\tilde{\mb{d}}_p| \lambda > \epsilon$ for $p\in\{\mu,\Delta\}$ where the reorganisation energy is
	\begin{equation}\label{eq:reorg}
		\lambda=\intf{0}{\infty}{\nu}\frac{J(\nu)}{\nu}=2\nu_cS.
	\end{equation}

	\section{Effects of strong permanent dipoles}\label{sec:ECF}
	In this section we derive the secularised PFME, provide a brief analytical review of it, and analyse the population transfer rates, decoherence rate, and Lamb shift appearing in the master equation, which are given by Fourier transforms of the environment correlation functions (ECFs). This will allow us to analytically explore the role of permanent dipoles. 
	
	Following the analytical discussion of the PFME, we will compare the PFME to the DFME, and to the numerically exact TEMPO \cite{strathearn2017efficient,Strathearn2018,pollock2018non,jorgensen2019exploiting,gribben2022exact,fux2021efficient,gribben2022using,fux2022thermalization,fowler2022efficient}, which both serve as benchmarks. In the numerical approach, we use the full non-secular Redfield master equations derived in Appendix~\ref{app:NS} for both frames. We emphasise that the non-secular PFME depends only on the physical processes discussed in the main text.
	
	In the eigenbasis, the secularised, time-local, Redfield master equation is
	\begin{subequations} 
		\label{eq:MEsec}
		\begin{align}
			\frac{\partial \rho_{++}(t) }{\partial t}&=-\gamma_\downarrow\rho_{++}(t)+\gamma_\uparrow\rho_{--}(t),\label{eq:MEseca}\\
			\frac{\partial \rho_{+-}(t) }{\partial t}&=-[\gamma_d+i\bar{\eta}]\rho_{+-}(t),
		\end{align}
	\end{subequations}
	and $(\partial /\partial t)\rho_{--}(t)=-(\partial /\partial t)\rho_{++}(t)$, $(\partial /\partial t)\rho_{-+}(t)=(\partial /\partial t)\rho_{+-}(t)^\dagger$, where $\rho_{ij}(t)=\braket{i|\rho_S(t)|j}$ for $i,j\in\{+,-\}$ and $\rho_S(t)=\Tr_E[\rho(t)]$. The transition and decoherence rates are
	\begin{subequations}\label{eq:ratesec}
		\begin{align}
			\gamma_{\substack{\downarrow\\\uparrow}}&=2\Re\left[\Gamma_{\mp\mp}(\pm \eta)\right],\label{eq:gr}\\
			\gamma_d&=\tfrac{1}{2}\left[\gamma_{\uparrow}+\gamma_{\downarrow}\right]+4\Re\left[\Gamma_{zz}(0)\right],\label{eq:gd}
		\end{align}
	\end{subequations}
	where\begin{align}\label{eq:Gamma}
		\Gamma_{\alpha\beta}(\omega)=\intf{0}{\infty}{s}\rme^{i\omega s}\left\langle g_{\alpha}^\dagger(s) g_{\beta}(0)\right\rangle,
	\end{align}
	for $\alpha,\beta\in\{z,+,-\}$ and $g_\alpha(s)$ is the interaction picture form of $g_\alpha$ in Eqs.~\eqref{eq:g}. Finally, the Lamb shifted eigenenergy is
	\begin{align}\label{eq:Lamb}
		\bar{\eta}=\eta+\Im\left[\Gamma_{--}(\eta)-\Gamma_{++}(-\eta)\right].
	\end{align}
	
	The secular master equation in Eqs.~\eqref{eq:MEsec} describes population transfer from $\ket{+}$ to $\ket{-}$ at decay rate $\gamma_{\downarrow}$, vice-versa at an excitation rate $\gamma_{\uparrow}$, decoherence at a rate $\gamma_d$, and oscillations in the coherence at frequency $\bar{\eta}$. 
	
	In the PFME we choose the initial state to be $\rho_p(0)=\proj{g}{g}\otimes \rho_E$ where $\rho_E$ is given in Eq.~\eqref{eq:rhoB} and assume that the environment state does not change throughout the dynamics. We discuss the implications of this initial state in Sec.~\ref{sec:IC}.

	Evaluating the ECFs in Eq.~\eqref{eq:Gamma}, $\langle g_{\alpha}^\dagger(s) g_{\beta}(0)\rangle$, is a laborious process and we present the full derivations in Appendix~\ref{app:gagb}. Here, we focus the discussion on new physical processes introduced by the presence of the strong permanent dipoles. The ECFs, $\langle g_{\alpha}^\dagger(s) g_{\beta}(0)\rangle$, depend on linear combinations of $\langle \mathcal{C}^\dagger(s)\mathcal{C}(0) \rangle$, $\langle \mathcal{C}(s)\mathcal{C}^\dagger(0)\rangle$, $\langle \mathcal{C}^\dagger(s)\mathcal{C}^\dagger(0)\rangle$ and $\langle \mathcal{C}(s)\mathcal{C}(0) \rangle$, with the relative weighting of each dependent on the eigenbasis angle $\varphi$. Importantly, as we show in Appendix~\ref{app:gagb}, each of these correlation functions describes the same physics. Therefore, for the analytics in the main text, we focus the discussion on 
 \begin{equation}
       \Gamma(\omega)=\intf{0}{\infty}{s}\rme^{i\omega s}\left\langle \mathcal{C}^\dagger(s) \mathcal{C}(0)\right\rangle.
\end{equation}
This function has contributions from four distinct processes,
 \begin{equation}\label{eq:g4}
\Gamma(\omega)=\Gamma_1(\omega)+\Gamma_2(\omega)+\Gamma_{V,1}(\omega)+\Gamma_{V,0}(\omega).
 \end{equation}
 The number in the subscripts of each term on the right-hand-side of Eq.~\eqref{eq:g4} denote how many photons are involved in the process (in the absence of the sideband), and terms with a subscript `$V$' are induced by the driving. The other three Fourier transforms of the $\mathcal{C}(s)$ two-time correlation functions also depend on these four contributions, except that $\langle \mathcal{C}^\dagger(s)\mathcal{C}^\dagger(0)\rangle$ and $\langle \mathcal{C} (s)\mathcal{C}(0)\rangle$ do not have a $\Gamma_{V,1}(\omega)$ type contribution. The prefactor of each term is dependent on the relative dipole angle $\theta_{\mu\Delta}$, the driving $|V|$, the relative phase $\vartheta_{\mu V}$, and dipole magnitudes $\Omega_{pq}$, as
 \begin{subequations}\label{eq:Gs}
	\begin{align}
		\Gamma_1(\omega)&\propto\Omega_{\mu\mu},\\
		\Gamma_2(\omega)&\propto\Omega_{\mu\Delta}^2\cos^2(\theta_{\mu\Delta}),\\
		\Gamma_{V,1}(\omega)&\propto\Omega_{\mu\Delta}\left|V\right|\cos\left(\vartheta_{\mu V}\right)\cos(\theta_{\mu\Delta}),\label{eq:GV1}\\
		\Gamma_{V,0}(\omega)&\propto\left|V\right|^2.
	\end{align}
 \end{subequations}
	From Eqs.~\eqref{eq:Gs} we see that when the system is driven, $\Gamma(\omega)$ is not an even function of $\theta_{\mu\Delta}$ or $\vartheta_{\mu V}$. We will later show that this can be utilised to control the system, for example to minimise decoherence. Moreover, for perpendicular dipoles $\cos(\theta_{\mu\Delta})=0$, so $\Gamma(\omega)$ becomes equivalent to the analogous function in the driven spin boson model where the transition dipole and permanent dipole interactions are with independent baths. Interactions proportional to $\cos(\theta_{\mu\Delta})$ are therefore arise because the interactions are non-commutative.
	
	The $\Gamma_{V,1}(\omega)$ contribution in Eq.~\eqref{eq:GV1} varies cosinusoidally with the relative complex phase $\vartheta_{\mu V}$. This means that through controlling the phase of $V$ one can determine the phase $\vartheta_\mu$ of $\mb{d}_\mu$, for example by minimising the spectral linewidth, allowing an estimation of the relative phase between the excited and ground state. This will work provided that $V$ is not generated via a coupling to the system's transition dipole moments, in which case $\vartheta_V=\vartheta_\mu+\vartheta_\text{external}$ and so $\vartheta_{\mu V}$ would only depend on the phase of the external drive.

 For additional context, when $|V|\ll \epsilon$ -- the parameter regime in which polaron theory is valid -- the eigenbasis angle is $\varphi\approx 0$ and so $g_z\approx0$ and $g_+\approx\mathcal{C}$. In this limit, the secular approximation used in the master equation in Eqs.~\eqref{eq:MEsec} becomes exact, and $\gamma_\uparrow\approx 2\Re[\Gamma(-\eta)]$ with $\Gamma(\omega)$ in Eq.~\eqref{eq:g4} and $\gamma_\downarrow\approx \int_0^\infty\text{d}s\ \rme^{i\eta s}\langle\mathcal{C}(s)\mathcal{C}^\dagger(0)\rangle$ which, as we show in Appendix~\ref{app:gagb}, evaluates to $\gamma_\downarrow\approx 2\Re[\bar{\Gamma}(\eta)]$ where $\bar{\Gamma}(\omega)$ is equal to Eq.~\eqref{eq:g4} but with $\Gamma_{V,1}(\omega)\to-\Gamma_{V,1}(\omega)$. (Note that this asymmetry cannot lead to population inversion due to minus signs within $\Gamma_{V,1}(\omega)$.)
	
	We will now discuss the four contributions to Eq.~\eqref{eq:g4} in turn. Each are derived in Appendix~\ref{app:gagb}.
	
	\subsection{One photon processes, $\Gamma_1(\omega)$}\label{sec:1pV0}
    The function describing driving-independent one photon processes is
	\begin{align}\label{eq:G1}
		\Gamma_1(\omega)=\pi\Omega_{\mu\mu}\intf{0}{\infty}{\nu}\big[&J_A(\nu)\mathcal{K}(\omega+\nu)\nonumber\\
  &\qquad+J_E(\nu)\mathcal{K}(\omega-\nu)\big],
	\end{align}
	where 
	\begin{equation}\label{eq:K}
		\mathcal{K}(\varepsilon)=\frac{1}{\pi}\intf{0}{\infty}{s}\rme^{i\varepsilon s}\rme^{\phi(s)-\phi(0)},
	\end{equation}
	contains the influence of the permanent dipoles within this term, and we have introduced the absorption and emission spectral densities,
	\begin{align}\label{eq:JAJE}
		J_A(\nu)&=J(\nu)N(\nu),\\
		J_E(\nu)&=J(\nu)\tilde{N}(\nu),
	\end{align}
	where $N(\nu)=1/[\exp(\beta\nu)-1]$ is the Bose-Einstein distribution and $\tilde{N}(\nu)=N(\nu)+1$ will be a useful notation throughout this paper.
 
	To understand the physical processes associated with this term,  it is convenient to temporarily ignore the factor of $\exp[\phi(s)-\phi(0)]$ in Eq.~\eqref{eq:K}. The resulting term is the typical function appearing in the standard optical master equation (SOME) \cite{breuer2002theory} - our model in the absence of permanent dipoles and driving  - which is
	\begin{align}\label{eq:SOME}
		&\Gamma_\text{SOME}(\omega)=\Omega_{\mu\mu}\bigg[\pi\left(J_A(-\omega)+J_E(\omega)\right)\nonumber\\
		&\qquad\qquad\qquad+i\mathcal{P}\intf{0}{\infty}{\nu}\left(\frac{J_A(\nu)}{\omega+\nu}+\frac{J_E(\nu)}{\omega-\nu}\right)\bigg],
	\end{align}
	where $\mathcal{P}$ denotes the principal value.
	Clearly, when $\omega=-\eta$, twice the real part of $\Gamma_\text{SOME}(\omega)$ describes excitation-by-absorption at a rate $2\pi\Omega_{\mu\mu} J_A(\eta)$, and when $\omega=\eta$ it describes decay-by-emission at a rate $2\pi\Omega_{\mu\mu} J_E(\eta)$. The imaginary part of $\Gamma_\text{SOME}(\omega)$ will determine the Lamb shift.
	
	We now return to Eq.~\eqref{eq:G1}. In Ref.~\cite{rouse2022analytic} an analytic solution to Eq.~\eqref{eq:K} was found by exploiting the fact that $\mathcal{K}(\varepsilon)$ only depends on $J(\nu)$ through its moments $\mu_m=4\Omega_{\Delta\Delta}\intf{0}{\infty}{\nu}[J(\nu)/\nu^2]\nu^m$ for $m=1,2,\ldots,\infty$, and that moments of lower order contribute relatively more. We can then evaluate $\mathcal{K}(\varepsilon)$ by replacing $J(\nu)$ with a truncated spectral density $J'(\nu)=\sum_{\mb{k}=1}^{N_*}|f'\uk|^2\delta(\nu-\nu\uk')$, as long as we choose the coupling strengths $\{f\uk'\}$ and energies $\{\nu\uk'\}$ of the modes such that $J'(\nu)/\nu^2$ has the same lowest moments as $\mu_m$ for $m=1,2,\ldots,2N_*$ \cite{rouse2022analytic}. The solution converges rapidly for increasing $N_*$, and a single-mode truncation is often very accurate, with truncation mode parameters: $\nu'_1\equiv\nu_s=\mu_2/\mu_1$ and $f'_1\equiv f_s=\mu_2^{1/2}\propto\Omega_{\Delta\Delta}^{1/2}$ \cite{rouse2022analytic}. A single-mode truncation captures all of the essential physics described by $\mathcal{K}(\omega)$. Thus, for the analytical analysis in this work, we use a single-mode truncation to evaluate Eq.~\eqref{eq:K}; however, in all of the simulations in this paper, we increase $N_*$ until convergence.
	
	From Ref.~\cite{rouse2022analytic}, the single mode solution of Eq.~\eqref{eq:K} is
	\begin{equation}\label{eq:Ksingle}
		\mathcal{K}(\varepsilon)=\sum_{\ell=-\infty}^\infty A_\ell\left[\delta(\varepsilon-\ell\nu_s)+\frac{i}{\pi}\frac{\mathcal{P}}{\varepsilon-\ell\nu_s}\right],
	\end{equation}
	where
	\begin{equation}\label{eq:Al}
		A_\ell=\sum_{n=|\ell|}^{\infty'}\sum_{m=\frac{n-\ell}{2}}^n\binom{n}{m}\binom{m}{m-\frac{n-\ell}{2}}W_nV_m,
	\end{equation}
	and the prime on the first summation indicates that only every other term is included, i.e. $n=|\ell|,|\ell|+2,\ldots,$ and $W_n=S^{ n}_s\exp[-S_s]/n!$ is the Franck-Condon factor of the mode in the truncation, $S_s=|f_s|^2/\nu_s^{2}=\mu_1^2/\mu_2$ is its Huang-Rhys factor and $V_m=N(\nu_s)^m\exp[-2S_sN(\nu_s)]$ \cite{rouse2022analytic}. $A_\ell$ has the normalisation property $\sum_{\ell=-\infty}^\infty A_\ell=1$, is maximised for $\ell=\text{Round}(S')$, becomes $A_\ell=\delta_{\ell 0}$ if $|\tilde{\mb{d}}_\Delta|=0$, and the $\ell<0$ terms are only non-zero at finite temperature (see Ref.~\cite{rouse2022analytic} for a full discussion on the mode truncation solution).
	
	Substituting Eq.~\eqref{eq:Ksingle} into Eq.~\eqref{eq:G1} yields
	\begin{equation}
		\Gamma_1(\omega)=\frac{1}{2}\gamma_1(\omega)+iS_1(\omega),
	\end{equation}
	where 
	\begin{equation}\label{eq:g0}
		\gamma_1(\omega)=2\pi\Omega_{\mu\mu}\sum_{\ell=-\infty}^\infty A_\ell\Big[J_A(\ell\nu_s-\omega)+J_E(\omega-\ell\nu_s)\Big],
	\end{equation}
	and 
	\begin{align}\label{eq:Ssb}
		S_1(\omega)=i\Omega_{\mu\mu}&\sum_{\ell=-\infty}^\infty A_\ell\mathcal{P}\intf{0}{\infty}{\nu}\nonumber\\
		&\times\bigg(\frac{J_A(\nu)}{\omega+\nu-\ell\nu_s}+\frac{J_E(\nu)}{\omega-\nu-\ell\nu_s}\bigg).
	\end{align}
	Eqs.~\eqref{eq:g0} and \eqref{eq:Ssb} show that the effect of the permanent dipoles within $\Gamma_1(\omega)$ is to introduce manifolds of harmonic levels (in the single mode truncation of $\mathcal{K}(\varepsilon)$ there is only one manifold) to/from which transitions can occur, and which influence the Lamb shift value. This is commonly observed when vibrational displacement interactions occur simultaneously with transition dipole interactions and manifest in spectra as vibrational side bands and Stokes's shift. However, in our model, this is a purely photonic effect and it results in a photon sideband in the spectrum of the system. $\Gamma_\text{SOME}(\omega)$ in Eq.~\eqref{eq:SOME} is recovered from $\Gamma_1(\omega)$ in the limit of no permanent dipoles because $A_{\ell}\to\delta_{\ell 0}$.
	
\begin{figure}[ht!]\centering
		\includegraphics[width=0.375\textwidth]{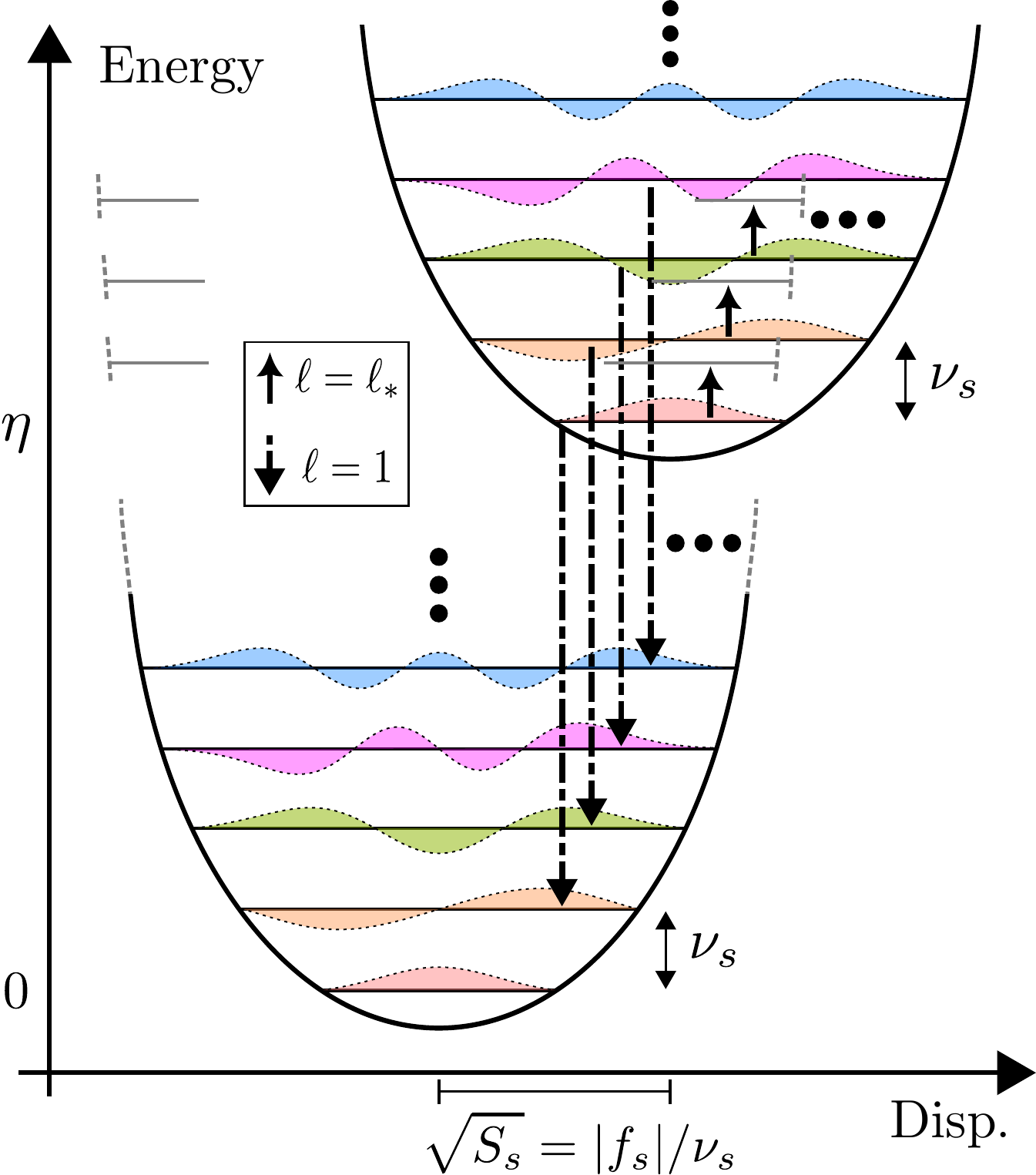}
		\caption{\textbf{Illustration of decay processes ($\omega=+\eta$) captured by the $\ell=1$ and $\ell=\ell_*$ terms in Eq.~\eqref{eq:g0}, within the single mode truncation of $\mathcal{K}(\varepsilon)$.} Decay-by-emission processes have dot-dashed arrows, and decay-by-absorption processes have solid arrows. The type of decay that occurs for a given value of $\ell$ depends on the sign of $\eta-\ell\nu_s$. In the absence of permanent dipoles $A_\ell=\delta_{\ell 0}$ and so only the $\ell=0$ decay processes are possible.}
		\label{fig:transitions}
	\end{figure}

	In Fig.~\ref{fig:transitions}, we illustrate decay-by-emission transitions corresponding to $\ell=1$, and decay-by-absorption transitions for $\ell=\ell_*$ where $\text{Round}(\eta/\nu_s)<\ell_*<\text{Round}(\eta/\nu_s)+1$. The $A_\ell$ values can be interpreted as the probabilities for a  decay between levels with energy difference $\eta-\ell\nu_s$ to occur. 
	
	\subsection{Two photon processes, $\Gamma_2(\omega)$}
	We now consider the two photon processes in Eq.~\eqref{eq:g4} that occur independent of the driving, $\Gamma_2(\omega)$. These processes require that the $\sigma_z$ and $\sigma_\pm$ do not commute, and so are unique to permanent dipole interactions. As shown in Appendix~\ref{app:gagb}, the rate function is
	\begin{align}\label{eq:G2}
		\Gamma_{2}(\omega)=&4\kappa^2\Omega_{\mu\Delta}^2\cos^2\left(\theta_{\mu\Delta}\right)\nonumber\\
&\times\sum_{n,m\in\{-1,1\}}\intf{0}{\infty}{s}\rme^{i\omega s+\phi(s)}\tilde{\chi}_{n,m}(s),
	\end{align}
	where 
	\begin{equation}\label{eq:Anmdagdot}
		\tilde{\chi}_{n,m}(s)=\intf{0}{\infty}{\nu}\intf{0}{\infty}{\nu'}\chi_{n,m}(\nu,\nu')\rme^{in\nu s}\rme^{im\nu's},
	\end{equation}
	is the inverse Fourier transform of $\chi_{n,m}(\nu,\nu')$ which we write in matrix notation as
	\begin{align}\label{eq:Adagdot}
		\chi(\nu,\nu')
		=\bar{J}(\nu)\bar{J}(\nu')\begin{pmatrix}
			\tilde{N}(\nu) \tilde{N}(\nu') & -\tilde{N}(\nu) N(\nu') \\
			-N(\nu)\tilde{N}(\nu') &  N(\nu) N(\nu')
		\end{pmatrix},
	\end{align}
	where, for example, $\chi_{1,1}(\nu,\nu')=4\Omega_{\mu\Delta}^2\bar{J}(\nu)\bar{J}(\nu')N(\nu)N(\nu')$. We have introduced the polarised spectral density,
	\begin{equation}\label{eq:Jbar}
		\bar{J}(\nu)=\frac{J(\nu)}{\nu},
	\end{equation}
	which will be relevant to all transitions resulting from permanent dipole interactions.
	
	The two-frequency Fourier transform in Eq.~\eqref{eq:Anmdagdot} indicates that $\Gamma_2(\omega)$ describes two simultaneous processes. From the phase factors in Eq.~\eqref{eq:Anmdagdot}, when $n$ or $m$ indices are equal to $-1$ we deal with emission processes, and when they are equal to $+1$ we deal with absorption, into the frequency channel $\nu$ and $\nu'$ for $n$ and $m$, respectively. This interpretation is reinforced by the positions of the factors of $N(\nu)$, $N(\nu')$, $\tilde{N}(\nu)$ and $\tilde{N}(\nu')$ in the matrix in Eq.~\eqref{eq:Adagdot}. Similar to the case of the one photon processes, the factor of $\exp[\phi(s)-\phi(0)]$ in Eq.~\eqref{eq:G2} will introduce a photon sideband to the overall process, again enabling processes with more photons.
	
\begin{figure}[ht!]\centering
		\includegraphics[width=0.48\textwidth]{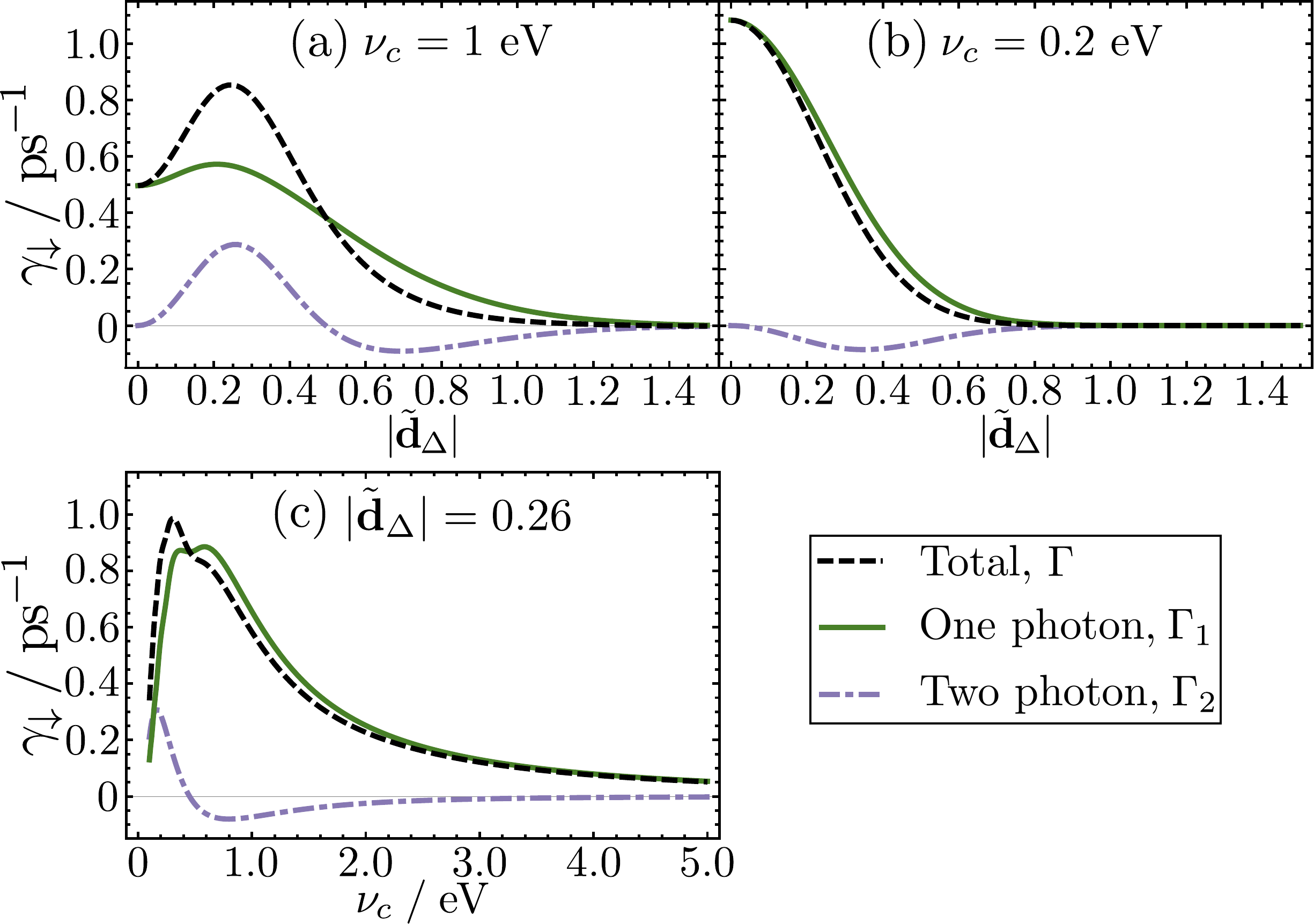}
		\caption{\textbf{Polaron frame decay rate for $V=0$.} (a) and (b) shows the total decay rate (dashed black), and the contributions of the one photon processes (solid green) and two photon processes (dot-dashed purple) verses $|\tilde{\mb{d}}_\Delta|$. In (a) the cutoff frequency is $\nu_c=0.2~\text{eV}$ and in (b) $\nu_c=1~\text{eV}$. (c) shows the total rate and its contributions verses $\nu_c$ for $|\tilde{\mb{d}}_\Delta|=0.26$ which corresponds to the maximum in (a).  Other parameters are $|\tilde{\mb{d}}_\mu|=0.01$, $\epsilon=1~\text{eV}$, $\beta=2~\text{eV}^{-1}$ ($T\approx 5800~\text{K}$), $S=1/\pi$, and $\theta_{\mu\Delta}=0$ which, unless otherwise stated, are used in all figures.}
		\label{fig:rates}
	\end{figure}

	Before we move onto the processes induced by driving, in Fig.~\ref{fig:rates} we plot the decay rate $\gamma_\downarrow$ as a function of $|\tilde{\mb{d}}_\Delta|$ and $\nu_c$ for $V=0$. In Fig.~\ref{fig:rates}, the eigenbasis coincides with the $\{\ket{e},\ket{g}\}$ basis ($\varphi=0$) and the decay rate is $\gamma_\downarrow=2\text{Re}[\Gamma(\epsilon)]$ where $\Gamma(\omega)$ is in Eq.~\eqref{eq:g4}. One can see that $\Gamma(\omega)$ is independent of the phase $\vartheta_\mu$ when $V=0$. 
 
 In Fig.~\ref{fig:rates}(a) and (b), it is clear that at small $\nu_c$ it is possible for permanent dipoles to enhance the decay rate, whilst at large $\nu_c$ they always lead to suppression. This is because of how the functions $J(\nu)$ and $\bar{J}(\nu)$ vary with $\nu$ at the sideband-modified frequencies compared with the changes in the mode populations governed by $N(\nu)$. Fig.~\ref{fig:rates}(c) shows that two photon processes become negligible for large $\nu_c$. This is because two photon processes scale as $J(\nu)J(\nu')\propto\nu_c^{-4}$ (see Eq.~\eqref{eq:SD}) whilst one photon processes scale as $J(\nu)\propto\nu_c^{-2}$. For the same reason, this will also occur in the limit of small Huang-Rhys parameter, $S\to 0$.
	
	\subsection{Driving induced one photon processes, $\Gamma_{V,1}(\omega)$}
	We now move onto the processes in Eq.~\eqref{eq:g4} induced by the driving. The function for such one photon processes is 
	\begin{align}\label{eq:GV1}
		\Gamma_{V,1}&(\omega)=4\pi\Omega_{\mu\Delta}|V|\cos\left(\vartheta_{\mu V}\right)\cos\left(\theta_{\mu\Delta}\right)\\
 & \times\intf{0}{\infty}{\nu}\bigg(-\bar{J}_A(\nu)\mathcal{K}(\omega+\nu)+\bar{J}_E(\nu)\mathcal{K}(\omega-\nu)\bigg)\nonumber,
	\end{align}
	where $\mathcal{K}(\varepsilon)$ is given in Eq.~\eqref{eq:K} and, analogously to Eqs.~\eqref{eq:JAJE}, we have introduced the polarised absorption and emission spectral densities,
	\begin{align}
		\bar{J}_A(\nu)&=\bar{J}(\nu)N(\nu),\\
		\bar{J}_E(\nu)&=\bar{J}(\nu)\tilde{N}(\nu),
	\end{align}
	where $\bar{J}(\nu)$ is in Eq.~\eqref{eq:Jbar}. Similarly to the driving-independent two photon processes, $\Gamma_2(\omega)$ in Eq.~\eqref{eq:G2}, these processes also arise due to the non-commutativity of the permanent and transition dipole interactions and vanish when $\mb{d}_\mu\cdot\mb{d}_\Delta=0$.
	
	Using the single mode truncation solution in Eq.~\eqref{eq:Ksingle}, we can decompose $\Gamma_{V,1}(\omega)$ into real and imaginary parts,
	\begin{equation}
		\Gamma_{V,1}(\omega)=\frac{1}{2}\gamma_{V,1}(\omega)+iS_{V,1}(\omega),
	\end{equation}
	which we do not write explicitly.
	Therefore, driving the system generates additional one photon (plus sideband photons) transitions, similar in nature to the transitions described by $\Gamma_1(\omega)$ but here scaling with $\Omega_{\mu\Delta}$ and dependent on the polarised spectral density. 
 
 Importantly, the prefactor in Eq.~\eqref{eq:GV1} is proportional to $\cos(\vartheta_{\mu V})\cos(\theta_{\mu\Delta})$ which allows one to control whether this function suppresses or enhances the rates and Lamb shifts. In practice, this control could be easy to achieve by tuning the driving phase.

	\subsection{Driving induced zero photon processes, $\Gamma_{V,0}(\omega)$}
	Finally, the zero photon processes induced by the driving in Eq.~\eqref{eq:g4} is 
	\begin{equation}\label{eq:GV0}
		\Gamma_{V,0}(\omega)=\kappa^2|V|^2 \intf{0}{\infty}{s}\rme^{i\omega s}\left(\rme^{\phi(s)}-1\right).
	\end{equation}
	This term vanishes when $\phi(s) = 0$ and so is generated entirely through the photon sideband. Eq.~\eqref{eq:GV0} also appears in the driven spin boson model describing phonon mediated transfer between eigenstates and so is an effect of pure dephasing.
	
\begin{figure}[ht!]\centering
		\includegraphics[width=0.48\textwidth]{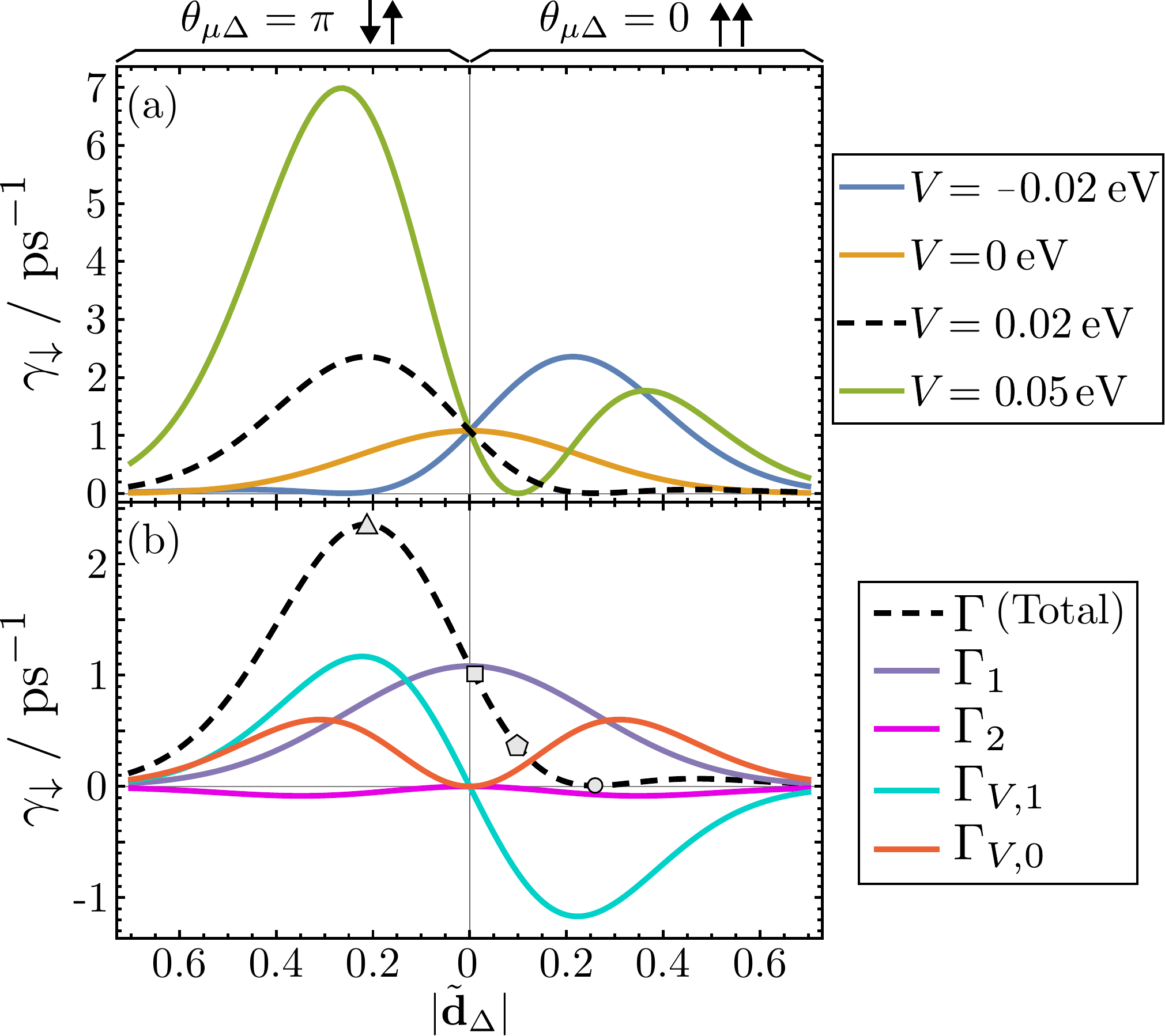}
		\caption{\textbf{Polaron frame decay rate for different permanent dipole and driving strengths.} (a) shows the total decay rate for different driving strengths. In all curves $\vartheta_\mu=0$ and $\vartheta_V\in\{0,\pi\}$ controls the sign of $V$ given in the legend. (b) shows the four contributions (coloured curves) to the total rate (dashed black), given in Eq.~\eqref{eq:g4}, for the $V=0.02~\text{eV}$ parameter set. The markers in panel (b) refer to $|\tilde{\mb{d}}_\Delta|$ values used in Fig.~\ref{fig:ratesangles}. The left-hand-side of each panel has anti-parallel $\mb{d}_\mu$ and $\mb{d}_\Delta$, whilst the right-hand-side has parallel dipole moments, which is indicated by the arrows at the top of the figure. $\nu_c=1~\text{eV}$, and other parameters are given at the end of the caption for Fig.~\ref{fig:rates}.}
		\label{fig:ratesVnot0}
	\end{figure}

	In Fig.~\ref{fig:ratesVnot0}, panel (a) shows the total decay rate for different driving strengths $V$ as a function of $|\tilde{\mb{d}}_\Delta|$, and panel (b) shows the four contributions to the rate with $V=0.02~\text{eV}$. The dashed black curves in both panels (a) and (b) are the same. The excitation $\gamma_{\uparrow}$ and decoherence $\gamma_d$ rates can be obtained from Fig.~\ref{fig:rates} by scaling the curves by $0.135$ and $0.568$, respectively; all three rates have the same dependence on the permanent dipoles unless $|V|\sim\epsilon$, but in that limit polaron theory becomes inaccurate. The key points resulting from the analysis of  Fig.~\ref{fig:rates} are: (1) one can both increase and decrease rates through the physical mechanisms enabled by the presence of permanent dipoles and driving, particularly by controlling $\Gamma_{V,1}(\omega)$ processes and (2) all physical processes found in the polaron ECFs contribute non-negligibly to the total rates. Recall from Fig.~\ref{fig:rates} that the contribution of the two photon processes $\Gamma_2(\omega)$ is larger for smaller $\nu_c$.

\begin{figure}[ht!]\centering
		\includegraphics[width=0.4\textwidth]{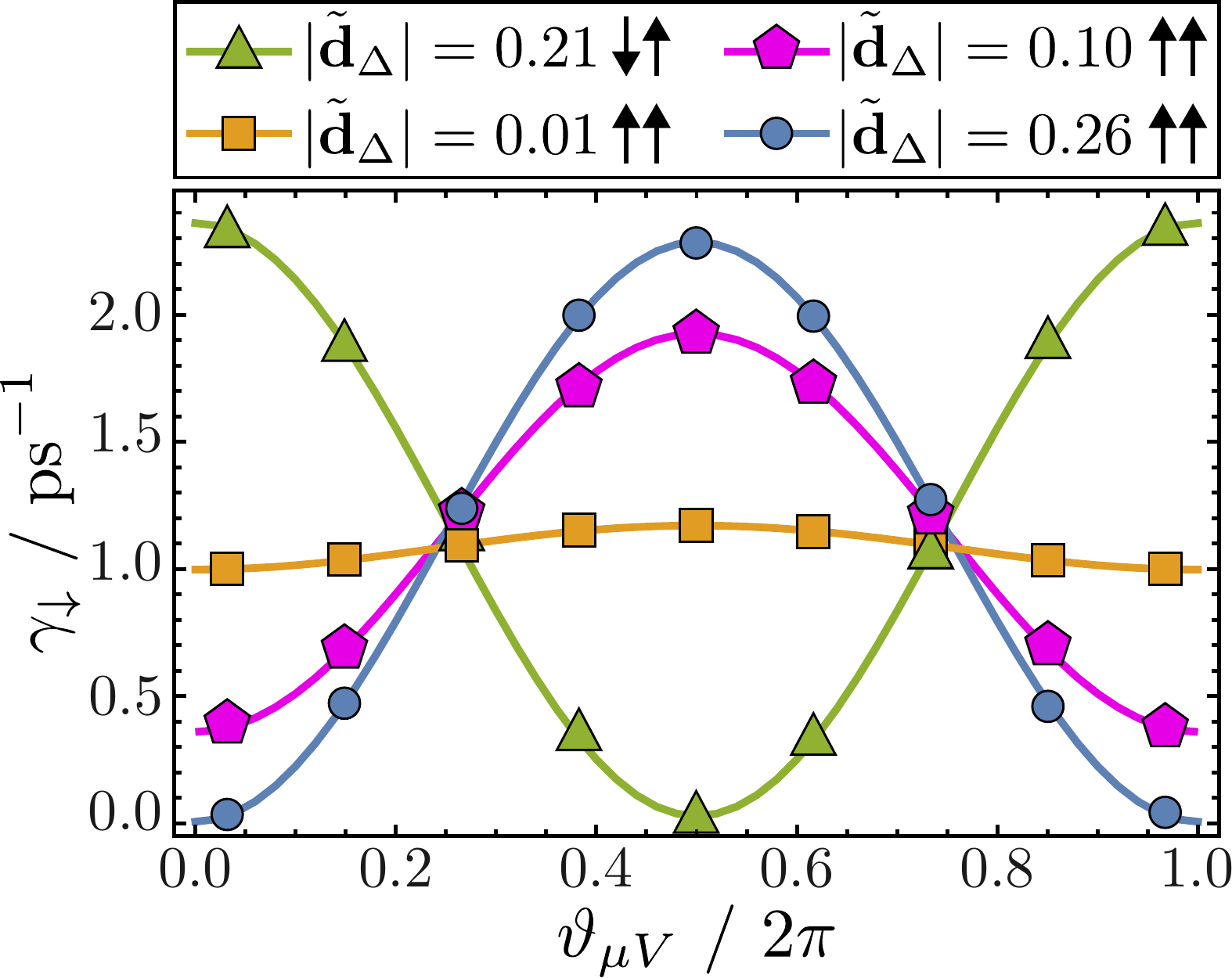}
		\caption{\textbf{Phase control of the polaron frame decay rate.} The relative phase $\vartheta_{\mu V}$ is varied whilst keeping $|\tilde{\mb{d}}_\mu|=0.01$ and $|V|=0.02~\text{eV}$ fixed. Each curve has a different value of $|\tilde{\mb{d}}_\Delta|$ indicated in the legend which correspond to the markers in Fig.~\ref{fig:ratesVnot0}(b). $\nu_c=1~\text{eV}$, and other parameters are given at the end of the caption for Fig.~\ref{fig:rates}.}
		\label{fig:ratesangles}
	\end{figure}

Through control of $|\tilde{\mb{d}}_\Delta|$ and $V$ one can suppress the transition and decoherence rates. For example, the decay rate at the minimum when $V=0.05$ in Fig.~\ref{fig:rates}(a), located at $|\tilde{\mb{d}}_\Delta|\approx 0.1$, is suppressed by a factor of $\approx 4000$ compared to its maximum value. Rather than tuning $|\tilde{\mb{d}}_\Delta|$, similar control can be achieved through the phase of the driving $V$. In Fig.~\ref{fig:ratesangles} we demonstrate this by varying $\vartheta_{\mu V}$ with $V=0.02$ and the $|\tilde{\mb{d}}_\Delta|$ values marked in Fig.~\ref{fig:ratesVnot0}(b). Maximum suppression is achieved when the driving is in, or out of, phase with the transition dipole moment, depending on the alignment of the dipole moments. This is because of the factor of $\cos(\theta_{\mu\Delta})\cos(\vartheta_{\mu V})$ in Eq.~\eqref{eq:GV1}.

	\section{Comparison to exact dynamics}\label{sec:ME}
\begin{figure*}[ht!]
		\centering
		\includegraphics[width=0.9\textwidth]{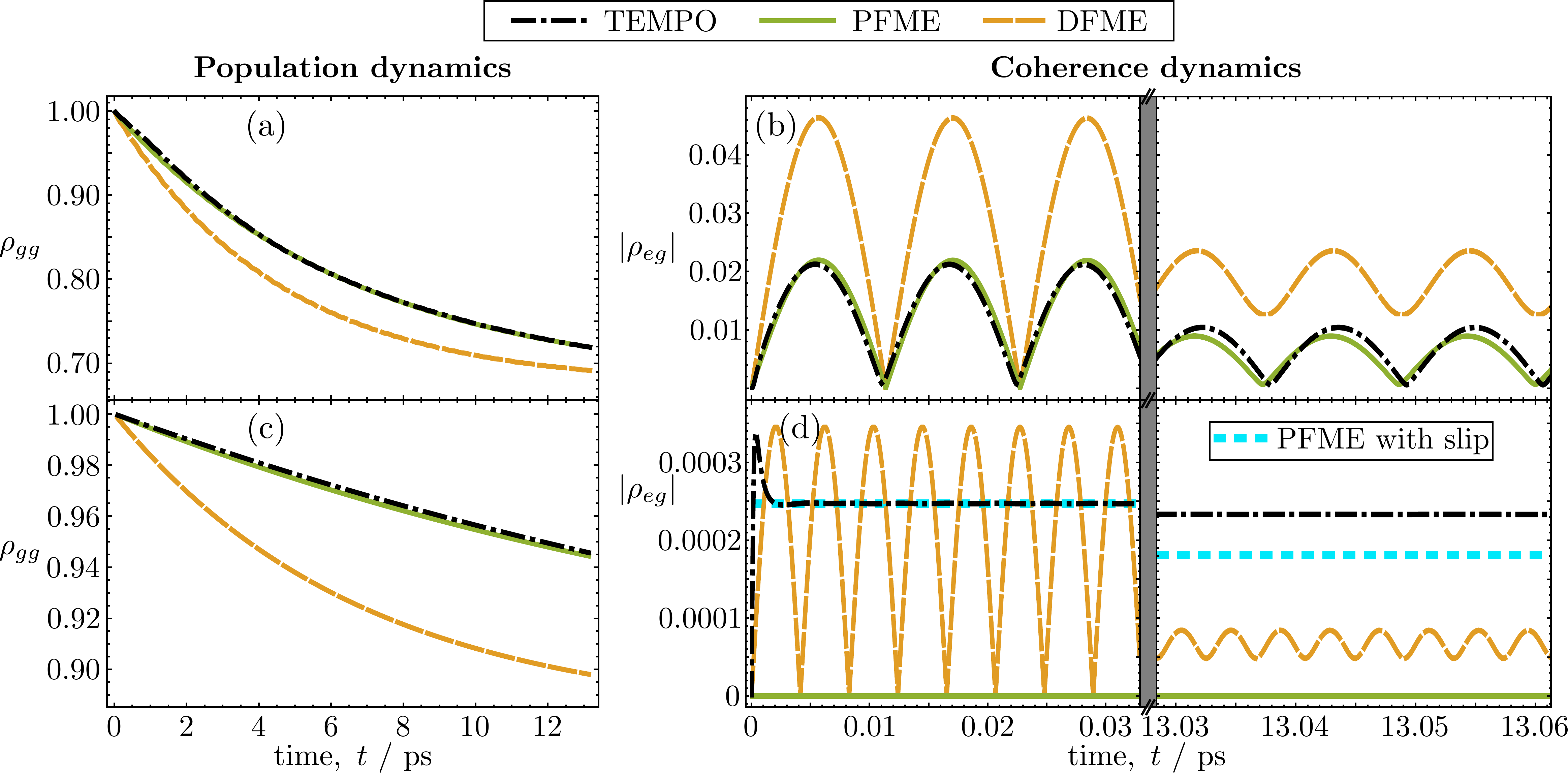}
		\caption{\textbf{Comparison of the PFME predictions to exact numerical approach and the DFME.} (a) and (c) are a comparison of $\rho_{gg}(t)$ calculated using the PFME (green) to TEMPO (dot-dashed black), and the DFME (dashed orange) over short and long times. (b) and (d) are a similar comparison but for $|\rho_{eg}(t)|$ (note the discontinuity in the time axis). In (d) the cyan dotted curve is the polaron prediction for the coherence artificially accounting for the initial non-Markovian slip. In (a) and (b) $\epsilon=0.36~\text{eV}$, $V=-0.0064~\text{eV}$, and $|\tilde{\mb{d}}_\Delta|=0.5$, and in (c) and (d) $\epsilon=1~\text{eV}$, $V=0~\text{eV}$, and $|\tilde{\mb{d}}_\Delta|=1$. In both plots $\nu_c=1~\text{eV}$, and other parameters are given at the end of the caption for Fig.~\ref{fig:rates}. To make comparison to TEMPO easier, we have assumed that the field aligns with the dipole moments, such that all dipoles are parallel and we have ignored the factor of $8\pi/3$ in the $\Omega_{pq}$ in Eqs.~\eqref{eq:gpq}. }
		\label{fig:MEcomparison}
	\end{figure*}

	In this section, we compare the predictions for the system density operator found using the PFME to those with the DFME and a numerically exact approach, TEMPO. The DFME does not capture many of the unique processes attributed to the permanent dipoles as seen in the polaron frame, and  we expect these perturbative approaches to become inaccurate when $|\tilde{\mb{d}}_\Delta|\lambda$ approaches $\epsilon$ where $\lambda=2\nu_c S$ is given in Eq.~\eqref{eq:reorg}. This is indeed manifest in Fig.~\ref{fig:MEcomparison} where we plot the population of the ground state and the coherence between the excited and ground state against time. 
	
	The PFME predicts qualitatively correct dynamics and agrees well with TEMPO, even for strong permanent dipoles. In Fig.~\ref{fig:MEcomparison}(a) and (b), $|\tilde{\mb{d}}_\Delta|=0.5$ and $\epsilon=0.36~\text{eV}$, and in (c) and (d) $|\tilde{\mb{d}}_\Delta|=1$ and $\epsilon=1~\text{eV}$, with $\lambda=2/\pi~\text{eV}$ in all panels. Hence, $|\tilde{\mb{d}}_\Delta|\lambda$ is $0.32~\text{eV}$ and $0.64~\text{eV}$ which is comparable to $\epsilon$ in both panels. This emphasises that each new process in Fig.~\ref{fig:rates} is required to correctly describe strong permanent dipoles.
	
	A phenomenon not captured by the PFME or any Markovian master equation is the so-called `slip' \cite{gaspard1999slippage} at short times, which is most evident in the coherence evolution in panel (d). However, as shown by the cyan dotted curve, if this is artificially accounted for in the PFME by starting in a coherent initial state, the PFME again agrees well over the long time duration indicating it captures the main decoherence mechanisms.

	\section{Emission spectrum}\label{sec:spectra}
	Due to the many new physical processes generated by the presence of permanent dipoles discussed in Sec.~\ref{sec:ECF} we expect alterations to the emission spectrum. In the absence of permanent dipoles, the spectrum consists of a Mollow triplet with peaks at frequencies $-\eta,0,\eta$. When $V=0$, the triplet becomes a single peak at frequency $\epsilon$ with a width determined by the decoherence rate in the standard optical master equation: $\pi J(\epsilon)[1+2N(\epsilon)]$. In the presence of strong permanent dipoles, we expect that there will be a photon sideband extending to negative frequencies. If $|\tilde{\mb{d}}_\Delta|$ is large enough, the positions of the Mollow triplet peaks will shift noticeably from $\pm\eta$ to $\pm\bar{\eta}$ given in Eq.~\eqref{eq:Lamb}, and the widths of the peaks will be determined by $\gamma_d$ in Eq.~\eqref{eq:gd}.
	
	As we prove in Appendix~\ref{app:spectrum}, the emission spectrum is given by
	\begin{equation}\label{eq:Semm}
		I(\omega)=\alpha_\text{prop}(\mb{r},\mb{R},\omega)I_0(\omega),
	\end{equation}
	where $\alpha_\text{prop}(\mb{r},\mb{R},\omega)=|\mb{d}_\mu\cdot\mb{G}(\mb{r},\mb{R},\omega)|^2$ and $\mb{G}(\mb{r},\mb{R},\omega)$ is the Green's function of the medium \cite{roy2015quantum,iles2017phonon,hughes2009theory,bundgaard2021non,walls1994gj}. $\alpha_\text{prop}(\mb{r},\mb{R},\omega)$ accounts for propagation and filtering of the light from the dipole to the detector at positions $\mb{r}$ and $\mb{R}$, respectively \cite{roy2015quantum}. The polarisation spectrum is
	\begin{equation}
		I_0(\omega)=\left\langle \sigma^+(\omega)\sigma^-(\omega)\right\rangle,
	\end{equation}
	or, equivalently,
	\begin{equation}\label{eq:S0}
		I_0(\omega)=\lim_{t\to\infty}\Re\left[\intf{0}{\infty}{\tau}\left\langle \sigma_+(t+\tau)\sigma_-(t)\right\rangle\rme^{-i\omega \tau}\right].
	\end{equation}
	In the following, we focus on evaluating  the polarisation spectrum, which captures entirely the effects associated with  the permanent dipoles. The specifics of the experimental set-up, for example, whether the dipole is coupled to a cavity or a waveguide, are described by the  $\alpha_\text{prop}(\mb{r},\mb{R},\omega)$ term, which can be calculated separately.
	
	The expectation value in Eq.~\eqref{eq:S0} is taken with respect to the lab frame density operator $\rho_l$. Using $\rho_l=U^\dagger \rho_p U$ where $U$ is the polaron transformation in Eq.~\eqref{eq:U} and $\rho_p$ is the polaron frame density operator, we have
	\begin{align}
		\left\langle \sigma_+(t+\tau)\sigma_-(t)\right\rangle&=\kappa^2\rme^{\phi(\tau)}\text{Tr}\left[ \sigma_+(t+\tau)\sigma_-(t)\rho_p(0)\right]\nonumber\\
		&\equiv\kappa^2\rme^{\phi(\tau)}\left\langle \sigma_+(t+\tau)\sigma_-(t)\right\rangle_p,
	\end{align}
	where $\kappa$ is defined above Eq.~\eqref{eq:phi}. Therefore, the final expression for the polaron frame polarisation spectrum is 
	\begin{align}\label{eq:S0p}
		I_{p,0}(\omega)=\kappa^2\lim_{t\to\infty}\Re\Big[&\intf{0}{\infty}{\tau}\rme^{\phi(\tau)}\nonumber\\
		&\times\left\langle \sigma_+(t+\tau)\sigma_-(t)\right\rangle_p\rme^{-i\omega \tau}\Big].
	\end{align}
	
	The two-time correlation function in Eq.~\eqref{eq:S0p} can be calculated using the quantum regression theorem (QRT) \cite{breuer2002theory,mccutcheon2016optical}. The QRT utilises the cyclicity of the trace to rewrite the two-time expectation value as
	\begin{equation}
		\lim_{t\to\infty}\left\langle \sigma_+(t+\tau)\sigma_-(t)\right\rangle_p=\text{Tr}\left[\sigma^+\Lambda_p(\tau)\right],
	\end{equation}
	where $\Lambda_p(\tau)=U_0^\dagger(\tau)\Lambda_p(0)U_0(\tau)$ is a modified density operator with the initial state \begin{equation}\label{eq:IC}
		\Lambda_p(0)=\sigma^-U_0(\infty)\rho_p(0)U_0^\dagger(\infty),
	\end{equation}
	and, because $U_0(\tau)$ is the time evolution operator defined by the polaron frame Hamiltonian (see Eq.~\eqref{eq:Hp}), the master equation we have derived holds identically for the $\Lambda_p(\tau)$ operator but with the initial condition defined by Eq.~\eqref{eq:IC}. The QRT allows one to convert two-time correlation functions into one-time expectation values of density operators with modified initial conditions.
	
	Due to the polaron transformation, Eq.~\eqref{eq:S0p} captures the photonic sideband. However, the QRT contains an implicit Born approximation \cite{mccutcheon2016optical} as well as any approximations used in the derivation of the master equation.
	
	Owing to the modified initial state in Eq.~\eqref{eq:IC}, the power of the polarisation spectrum contains information about the steady state of the system,
	\begin{equation}\label{eq:P}
		P=\intf{-\infty}{\infty}{\omega}I_{p,0}(\omega)=\pi \rho_{ee}(\infty).
	\end{equation}
	Additionally, by noting that the polarisation spectrum without the photon sideband, denoted by $I_{p,\times}(\omega)$, is given by Eq.~\eqref{eq:S0p} with the replacement $\exp[\phi(\tau)]\to1$, one can show that
	\begin{equation}\label{eq:SP}
		\frac{\intf{-\infty}{\infty}{\omega}I_{p,\times}(\omega)}{P}=\kappa^2.
	\end{equation}
	Hence, $\kappa^2$ can be interpreted as the fractional emission into the sideband. Remarkably, an integrated spectrum measurement can be used to determine $\kappa$, which in turn provides an effective measurement of $\Omega_{\Delta\Delta}$ and, therefore, of the strength of the $|\tilde{\mb{d}}_\Delta|$ dipole moment. In measurements, $I_{p,\times}(\omega)$ cannot be separated from $I_{p,0}(\omega)$, but as we show in the following, $\intf{-\infty}{\infty}{\omega}I_{p,\times}(\omega)$ can be accurately approximated as the integral of the measured spectrum over the Lorentzian parts of the dominant peaks \cite{trebbia2022tailoring}.

 We note that using Eq.~\eqref{eq:P} to determine $\rho_{ee}(\infty)$ requires the spectrum to be measured in units of inverse energy such that $P$ is dimensionless. This may be difficult for experiments that measure the spectrum in, for example, Watts per area or counts of photodetection events. However, the measurement of $\kappa^2$ in Eq.~\eqref{eq:SP} is independent of the units used to measure the spectrum.
	
\begin{figure}[ht!]\centering
		\includegraphics[width=0.48\textwidth]{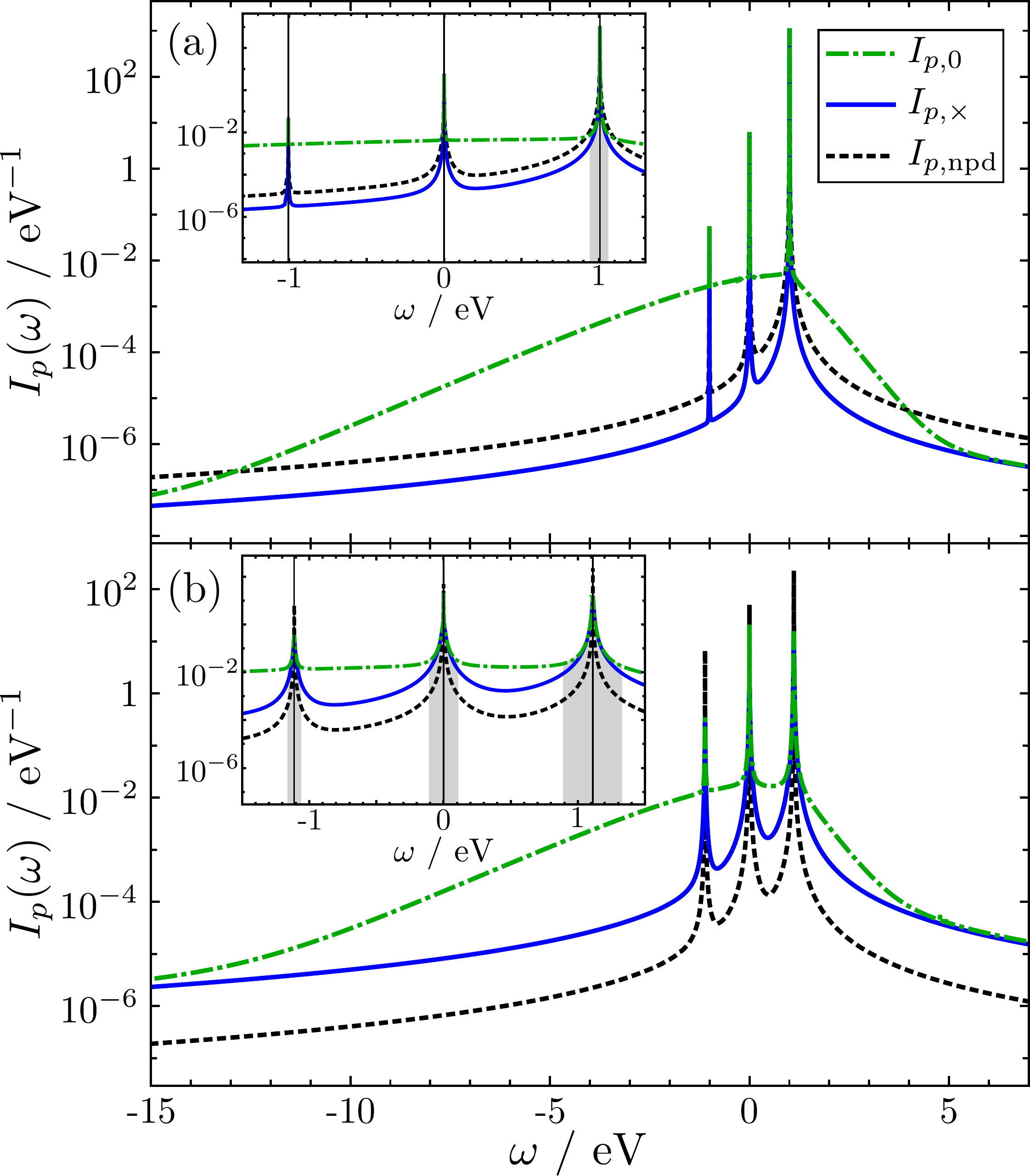}
		\caption{\textbf{Polarisation spectrum of the system.} The total polarisation $I_{p,0}(\omega)$ is shown in dot-dashed green, the polarisation spectrum without the photon sideband $I_{p,\times}(\omega)$ in solid blue, and the polarisation spectrum with $|\mb{d}_\Delta|=0$ (and other parameters unchanged) $I_{p,\text{npd}}(\omega)$ is shown in dashed black. In (a) $|\tilde{\mb{d}}_\Delta|=0.05$ and $V=0.05~\text{eV}$ and in (b) $|\tilde{\mb{d}}_\Delta|=0.1$ and $|V|=0.25~\text{eV}$. Other parameters are given in the caption at the end of Fig.~\ref{fig:rates}. The insets show shaded regions for which $\intf{-\infty}{\infty}{\omega}I_{p,\times}(\omega)\approx\int_R\text{d}\omega\ I_{p,0}(\omega)$ where $R$ is the domain of the shaded regions, which can be used in Eq.~\eqref{eq:SP} to estimate $\kappa$ from a measured spectrum. We use the secularised master equation to produce the spectra, which eliminates a known inconsistency near to the negative frequency peak and has no other impact.}
		\label{fig:spectra}
	\end{figure}

	In Fig.~\ref{fig:spectra}, we plot the polarisation spectrum calculated in the polaron frame [$I_{p,0}(\omega)$] in dot-dashed green, in blue we plot the spectrum for the same parameters but without the photonic sideband [$I_{p,\times}(\omega)]$ which serves to highlight the sideband, and in dashed black we plot the polarisation spectrum in the absence of permanent dipoles [$I_{p,\text{npd}}(\omega)$]. In panel (a) $|\tilde{\mb{d}}_\Delta|=0.05$ and $V=0.05~\text{eV}$, and in (b) $|\tilde{\mb{d}}_\Delta|=0.1$ and $V=0.25~\text{eV}$. Notably, the photon sideband substantially changes the emission spectrum even for small $|\tilde{\mb{d}}_\Delta|$. The widths of the peaks are noticeably larger in panel (b) due to the permanent dipoles; however, the effect of the permanent dipoles on the Lamb shifts is negligible.

	We now introduce a procedure of directly obtaining essential information about the system from the experimentally measurable spectrum, $I_{p,0}(\omega)$, and compare the values obtained for the data shown in Fig.~\ref{fig:spectra} to analytic values. We show that it is possible to evaluate $\rho_{ee}(\infty)$ and $\kappa$ from $I_{p,0}(\omega)$ using Eqs.~\eqref{eq:P} and \eqref{eq:SP} and, provided that the temperature of the experiment is known, these values can be used to obtain $\epsilon$ and $|V|$. If the spectral density is also known, it is also possible to obtain $|\tilde{\mb{d}}_\Delta|$ from $\kappa$. The values of the parameters obtained from the spectrum (methods explained afterwards) and the analytic values are given in Table~\ref{tab:values}.
	
	\begin{table}[h!]
		\centering
		\begin{tabular}{ c | c c | c c } 
			\hline\hline
			Param. & Analytic (a) & Meas. (a) & Analytic (b) & Meas. (b)\\
			$\kappa^2$  & 0.962 & 0.953 & 0.855 & 0.832 \\
			$\rho_{ee}(\infty)$ & 0.120 & 0.119 & 0.136 & 0.113\\
			$\epsilon/\text{eV}$ & 1.000 & 1.004 & 1.000 & 1.071\\
			$|V|/\text{eV}$ & 0.050 & 0.027 & 0.250 & 0.164\\
			$|\tilde{\mb{d}}_\Delta|$ & 0.050 & 0.050 & 0.100 &0.108\\
			\hline\hline
		\end{tabular}
		\caption{\textbf{Values of parameters obtainable from the measurable spectrum.}  Each column shows analytic values, obtained directly from the equations in this paper, and `measured' values obtained from $I_{p,0}(\omega)$ in Fig~\ref{fig:spectra}. The left-hand column is for Fig.~\ref{fig:spectra}(a) and the right-hand column for Fig.~\ref{fig:spectra}(b).}
		\label{tab:values}
	\end{table}
	
	As shown in Eq.~\eqref{eq:P}, the steady state population of the excited state can straightforwardly be obtained by integrating $I_{p,0}(\omega)$ over all frequencies. $\kappa$ can be approximately found using Eq.~\eqref{eq:SP} with $\intf{-\infty}{\infty}{\omega}I_{p,\times}(\omega)\approx\int_R\text{d}\omega\ I_{p,0}(\omega)$ where $R$ are the frequency regions for which the dominant peaks in the spectrum are approximately Lorentzian before the photon sideband begins. These regions are shown by the shaded regions in the insets of Fig.~\ref{fig:spectra}. In principle, one could increase the accuracy of the $\kappa^2$ estimation in Fig.~\ref{fig:spectra}(a) by integrating over the other two peaks, but these contributions are small. If the spectral density of the photon bath is known, the relation $\kappa=\exp[-\phi(0)/2]$ can be inverted to obtain $|\tilde{\mb{d}}_\Delta|$. The remaining parameters of the system, $\epsilon$ and $|V|$, can be estimated by assuming that the system thermalises with respect to $H_{S}=\eta\tau_z/2$, yielding the steady state
	\begin{equation}\label{eq:rhoee}
		\rho_{ee}(\infty)=\frac{1}{2}\left[1-\frac{\epsilon}{\eta}\tanh\left(\frac{\beta\eta}{2}\right)\right].
	\end{equation}
	Using Eq.~\eqref{eq:rhoee}, the measured value of $\rho_{ee}(\infty)$, and assuming negligible Lamb shifts such that the frequency of the peak in the spectrum $\bar{\eta}\approx\eta$, one can estimate $\epsilon$. Then, using $\bar{\eta}\approx\eta=(\epsilon^2+4\kappa^2 |V|^2)^{1/2}$, one can estimate $|V|$. The estimations of $|V|$ in Table~\ref{tab:values} are the least accurate owing to the sensitivity of both Eq.~\eqref{eq:rhoee} and $\eta$ to inaccuracies in $\rho_{ee}(\infty)$ and $\epsilon$, respectively.

	\section{Importance of initial conditions}\label{sec:IC}
	 It is typical in theoretical quantum optics to consider the system and environment initially in an uncorrelated state,
	\begin{equation}
		\rho(0) = \rho_S(0)\otimes\rho_E,
		\label{eqn:InitCondSep}
	\end{equation}
	where the environment is assumed to be in a thermal Gibbs state $\rho_E = \exp(-\beta H_E)/\mathcal{Z}_E$ with $H_E=\sum_k\nu\uk a_k^\dagger a_k$. This assumption is valid in weakly coupled systems where the system and the environment are only weakly correlated even after thermalisation. However, for strongly coupled systems, this assumption is invalidated, and the total Gibbs state $\rho_\beta=\exp(-\beta H)/\mathcal{Z}$ should be used as the initial state where $H$ now refers to the total Hamiltonian. Due to the interaction term, this will no longer be a separable state and so is difficult to model.
	
	A benefit of the polaron transformation is that the separable initial state \mbox{$\rho_S(0)\otimes\rho_E$} in the polaron frame, models an initial state in the lab frame that is more similar to $\rho_\beta$. In our calculations, $\rho_S(0)=\proj{g}{g}$, and so $\rho_S(0)\otimes\rho_E$ in the polaron frame becomes $\rho_l(0)=\proj{g}{g}\otimes B([D+\Delta]/\nu) \rho_E B(-[D+\Delta]/\nu)$ in the lab frame, which is equal to $\rho_\beta$ in the limit of negligible transition dipoles and projected onto the ground state. This is the state of the system after decaying to the ground state from a thermalised state, and so is a good initial state in which to model excitation.
	
	Therefore, to compare the PFME to TEMPO (which operates in the lab frame), we must use the Hamiltonian in Eq.~\eqref{eq:Hl} with the initial state $\rho_l(0)$. However, most numerical techniques, including TEMPO, assume that the initial state is $\rho_S(0)\otimes\rho_E$. In Appendix~\ref{app:EffHam}, we show that this discrepancy can be overcome by deriving an effective Hamiltonian to use in TEMPO such that the environment appears to be in the correct state. For $\rho_S(0)=\proj{g}{g}$ in the polaron frame, the required Hamiltonian is $\bar{H}=\bar{H}_0+\bar{H}_I$,
	where
	\begin{subequations}\label{eq:EffHam2}
		\begin{align}
			\bar{H}_0&=\frac{\tilde{\epsilon}}{2}\sigma_z+\tilde{V}\sigma_++\tilde{V}^*\sigma_-+\sum_k\nu\uk a_k^\dagger a_k,\\
			\bar{H}_I&=A_{\Delta\Delta} \left(\mathcal{I}+\sigma^z\right)+\pi_{\mu\bar{\mu}}\sigma_++\pi_{\bar{\mu}\mu}\sigma_-,
		\end{align}
	\end{subequations}
	and $\pi_{pq}$ is given in Eq.~\eqref{eq:Apq}, $\tilde{\epsilon}=\epsilon+2G_{\Delta\Delta}$, $\tilde{V}=V+G_{\mu\bar{\mu}}$, where $G_{pq}=\sum_k(p_k\Delta_k^*+q_k^*\Delta_k)/\nu\uk$. We note that because the displacement direction of the polaron transformation is state dependent, the necessary effective Hamiltonian depends on $\rho_S(0)$.
	
	By comparing the Hamiltonians in Eqs.~\eqref{eq:EffHam2} and Eq.~\eqref{eq:Hl}, we see stark differences if both are assumed to have the initial state $\rho_S(0)\otimes\rho_E$. This includes a renormalisation of the transition energy $\epsilon\to\tilde{\epsilon}$ and the driving term $V\to\tilde{V}$. 
	We can understand these differences by noting that the environment is far from equilibrium for a strongly coupled system, and initially, no optical polarons exist in our system. Dynamically created optical polarons effectively introduce a strong restoring force in the system that scales with the strength of the permanent dipoles. The ensuing dynamics are thus considerably different, exacerbating the need for careful consideration of the initial conditions of the physical models we deploy.

Identity type interactions, such as $\pi_{DD}\mathcal{I}'$ in Eq.~\eqref{eq:Hl}, have a similar effect. As we showed in Eq.~\eqref{eq:Hd}, this interaction can be removed through a displacement transformation and so, ultimately, the role of an identity interaction is to change the initial environment state from a thermal Gibbs state to a displaced one.

	\section{Conclusion}
	\label{sec:conc}
	We have studied a driven quantum optical system with strong permanent dipole moments associated with molecular orbital asymmetry. The optical polaron transformation, which captures the polarisation of photonic modes caused by the permanent dipoles, allows the construction of an optical master equation perturbative only in the transition dipole moment and driving strength, and provides an intuitive formalism to understand the effects associated with the presence of the permanent dipoles on the system dynamics and emission spectrum.
	
	We have shown three key results. (1) In Sec.~\ref{sec:ECF}, we showed that transition and decoherence rates can be engineered for practical application by exploiting permanent dipoles, for example, to reduce decoherence. By using the novel physical processes explicit in the polaron rate equations we derive, one can design systems to exploit these effects. The novel physical processes arising from the permanent dipole are: a photon sideband with a relative contribution to the emission spectrum scaling as $\kappa^2$, two-photon processes scaling as $\Omega_{\mu \Delta}^2$, and a term linear in the driving amplitude. The linear driving term is particularly useful for engineering enhanced or suppressed rates by varying the driving phase. (2) In Sec.~\ref{sec:ME}, we proved that the optical polaron description provides a much more accurate master equation by comparing it to the DFME and TEMPO. (3) In Sec.~\ref{sec:spectra}, we indicated distinguishable features of permanent dipoles in emission spectra and described possible measurements to obtain $\rho_{ee}(\infty)$, $\kappa$, $\bar{\eta}$ and by extension the bare energy splitting $\epsilon$ and the permanent dipole magnitude $|\tilde{\mb{d}}_\Delta|$. The distinguishable features are the photonic sideband and the altered width and position of peaks.

Furthermore, two of the processes we identified, namely the $\Gamma_2(\omega)$ and $\Gamma_{V,1}(\omega)$ processes, originate from the non-commutativity of the bosonic operators on the transition and permanent dipole interactions. These vanish if $\mb{d}_\mu$ and $\mb{d}_\Delta$ are perpendicular in which case our model becomes the driven spin boson model. We have also shown that in the absence of driving, this model maps onto the independent boson model in the limit of large cut off frequency ($\nu_c\to\infty$) or small Huang-Rhys parameter ($S\to 0$) where two photon processes are suppressed by the prefactor of the spectral density in Eq.~\eqref{eq:SD}.
	
	There are many interesting features of optical polarons that warrant future exploration. For example, how the transition rates are affected by embedding the dipole in a structured \cite{burgess2022dynamical}, or anisotropic \cite{messinger2020spontaneous}, dielectric medium, common to many biological systems. Moreover, many asymmetric systems couple strongly to vibrational baths and the interplay between photonic and vibrational physics leads to non-additive and non-equilibrium phenomena such as population inversion \cite{gribben2022exact,maguire2019environmental}. How permanent dipoles affect these phenomena is still an open question. Additionally, it has been shown in Refs.~\cite{guarnieri2018steady,  roman2021enhanced,purkayastha2020tunable} that the interplay between the pure dephasing and dissipative interactions leads to non-zero coherences in the steady state. It would be of great interest to explore  the nature of the steady state coherences in the optical polaron formalism.

	\section*{Acknowledgements}
	We would like to thank the Tempo Collaboration for use of the open-source code Oqupy \cite{oqupy}. In particular, we thank Gerald Fux for extremely insightful conversations on the use of Oqupy. D.M.R. also thanks Ahsan Nazir and Owen Diba for helpful discussions. 
	
	The work by A.B. was supported by the Leverhulme Quantum Biology Doctoral Training Centre at the University of Surrey funded by a Leverhulme Trust training centre grant number DS-2017-079, and the EPSRC (United Kingdom) Strategic Equipment Grant No. EP/L02263X/1 (EP/M008576/1) and EPSRC (United Kingdom) Grant EP/M027791/1 awards to M.F. D.M.R. is supported by EPSRC (United Kingdom) Grant EP/T517896/1.

	\bibliographystyle{PRX}	 
	\bibliography{bibPD}
	
	\clearpage
	\onecolumngrid
	\appendix


	\section{Displacement operators}\label{app:disp}
	Within the appendices, we will regularly use many identities involving displacement operators which are proven in Ref.~\cite{nazir2016modelling}. This first appendix is dedicated to listing the necessary displacement operator identities. For the purposes of this appendix, we use displacement operators with a single photon mode 
	\begin{equation}\label{eq:D1}
		B(\alpha)=\exp[\alpha a^\dagger - \alpha^* a],
	\end{equation}
	and note that $B(\alpha)^\dagger=B(-\alpha)$ and $B(-\alpha)B(\alpha)=\mathcal{I}$. The first identity is the action of a displacement operator on harmonic operators,
	\begin{subequations}\label{eq:D2}
		\begin{align}
			B(\pm\alpha) a^\dagger B(\mp\alpha)&=a^\dagger\mp\alpha^*,\\ B(\pm\alpha) a B(\mp\alpha)&=a\mp\alpha.
		\end{align}	
	\end{subequations}
	The second is that the product of two displacement operators is
	\begin{equation}\label{eq:D3}
		B(\alpha_1)B(\alpha_2)=B(\alpha_1+\alpha_2)\rme^{\frac{1}{2}\left(\alpha_1\alpha_2^*-\alpha_1^*\alpha_2\right)}.
	\end{equation}
	The third is the action of a displacement operator on the vacuum state to generate a coherent state,
	\begin{equation}\label{eq:D4}
		B(\alpha)\ket{0}=\ket{\alpha}.
	\end{equation}
	Fourth, that the expectation value of an operator with respect to the thermal state $\rho_E=\exp[-\beta\nu a^\dagger a]/Z_E$ can be written as an integral over coherent states as
	\begin{equation}\label{eq:D5}
		\Tr_E[O\rho_E]=\frac{1}{\pi N(\nu)}\int_{-\infty}^{\infty}d^2\alpha\ \rme^{-\frac{\left|\alpha\right|^2}{N(\nu)}}\braket{\alpha\left|O\right|\alpha},
	\end{equation}
	where $\int_{-\infty}^{\infty}d^2\alpha=\int_{-\infty}^{\infty}d\Im[\alpha]\int_{-\infty}^{\infty}d\Re[\alpha]$ and $N(\nu)=(\rme^{\beta\nu}-1)^{-1}$ is the Bose-Einstein distribution. The final identity is the expectation value of a displacement operator with respect to the vacuum state,
	\begin{equation}\label{eq:D6}
		\braket{0\left|B(\alpha)\right|0}=\rme^{-\frac{1}{2}\left|\alpha\right|^2}.
	\end{equation}

	\clearpage

	\section{Calculation of $\left\langle C\right\rangle$}\label{app:Cexp}
	Recall from Eq.~\eqref{eq:C} that $C=B(\delta)A_{\mu\bar{\mu}}B(\delta)+VB(2\delta)$ where $\delta_k=\Delta_k/\nu\uk$. Using properties of displacement operators this can be written as
	\begin{equation}\label{eq:Capp}
		C=B(2\delta)\left(\sum_k\left[\mu_k\left(a_k^\dagger+\delta_k^*\right)+\bar{\mu}_k^*\left(a_k+\delta_k\right)\right]+|V|\right).
	\end{equation}
	In order to calculate $\langle C\rangle=\Tr_E[C\rho_E]$ we require $\langle B(2\delta)\rangle$, $\langle B(2\delta)a^\dagger_k\rangle$ and $\langle B(2\delta)a_k\rangle$. We will perform these calculations explicitly, because the polaron frame environment correlation functions require analogous, but more algebraically involved, mathematics. We do the calculations with a single photonic mode, where $B(\alpha)=\exp[\alpha a^\dagger-\alpha a]$, and reinstate the multimode summations at the end. To calculate the expectation values we aim to use Eq.~\eqref{eq:D5}.
	
	Starting with $\langle B(2\delta)\rangle$, we first evaluate the integrand of Eq.~\eqref{eq:D5} as
	\begin{align}\label{eq:Dpm2hEX}
		\braket{\alpha\left|B(2\delta)\right|\alpha}&=\braket{0\left|B(-\alpha)B(2\delta)B(\alpha)\right|0}\nonumber\\
		&=\braket{0\left|B(2\delta)\right|0}\rme^{2\delta\alpha^*-2\delta^*\alpha}\nonumber\\
		&=\rme^{-2\left|\delta\right|^2}\rme^{2\delta\alpha^*-2\delta^*\alpha},
	\end{align}
	where in the first equality we have used Eq.~\eqref{eq:D4} twice, in the second equality Eq.~\eqref{eq:D3} twice and in the final equality Eq.~\eqref{eq:D6}. Substituting Eq.~\eqref{eq:Dpm2hEX} into Eq.~\eqref{eq:D4} then yields
	\begin{equation}\label{eq:C1}
		\left\langle B(2\delta)\right\rangle=\rme^{-2\left|\delta\right|^2\coth\left(\frac{\beta\nu}{2}\right)}\equiv\kappa.
	\end{equation}
	
	Evaluating $\langle B(2\delta)a^\dagger\rangle$ follows a similar procedure. However one must now use Eqs.~\eqref{eq:D2} to move the creation operator such that it annihilates with the vacuum bra-state $\bra{0}$. That is,
	\begin{align}\label{eq:Dpmcd2hEX}
		\braket{\alpha\left|B(2\delta)a^\dagger\right|\alpha}&=\braket{0\left|B(-\alpha)B(2\delta)a^\dagger \right|\alpha}\nonumber\\
		&=\braket{0\left|B(-\alpha)\left[a^\dagger- 2\delta^*\right]B(2\delta)\right|\alpha}\nonumber\\
		&=\left[\alpha^*- 2\delta^*\right]\braket{\alpha \left|B(2\delta)\right|\alpha}\nonumber\\
		&=\left[\alpha^*- 2\delta^*\right]\rme^{-2\left|\delta\right|^2}\rme^{2\delta\alpha^*-2\delta^*\alpha}.
	\end{align}
	Substituting Eq.~\eqref{eq:Dpmcd2hEX} into Eq.~\eqref{eq:D5} and performing the integrations yields
	\begin{equation}\label{eq:C2}
		\langle B(2\delta)a^\dagger\rangle=-2\kappa\delta^*\left[1+N(\nu)\right].
	\end{equation}
	The final expectation value, $\langle B(2\delta)a\rangle$, is easier to calculate because $a$ annihilates with $\ket{0}$. One finds that
	\begin{equation}
		\braket{\alpha\left|B(2\delta)a\right|\alpha}=\alpha\rme^{-2\left|\delta\right|^2}\rme^{2\delta\alpha^*-2\delta^*\alpha},
	\end{equation}
	and so
	\begin{equation}\label{eq:C3}
		\left\langle B(2\delta)a\right\rangle=2\kappa\delta N(\nu).
	\end{equation}
	
	Collecting the expectation values in Eqs.~\eqref{eq:C1}, \eqref{eq:C2} and \eqref{eq:C3} and substituting these into $\langle C\rangle$ with $C$ given in Eq.~\eqref{eq:Capp} yields
	\begin{equation}
		\langle C\rangle=\kappa\left[\coth\left(\frac{\beta\nu}{2}\right)\left(\bar{\mu}^*\delta-\mu\delta^*\right)+|V|\right].
	\end{equation}
	Since $\mb{d}_\Delta \in \Re$, we find that $\bar{\mu}^*\delta-\mu\delta^*=0$, and so $\langle C\rangle =\kappa |V|$ as stated in the main text.

	\clearpage

	\section{Non-secular master equations}\label{app:NS}
	The non-secular master equation in both the polaron and displaced frames have the same forms, only differing in which operators enter the two-time correlation functions, and therefore the rates and energies in the master equation below will be different in either. We find that
	\begin{subequations} \label{eq:nonsec}
		\begin{align}
			\partial_t\rho_{++}(t)&=-\gamma_{\downarrow}\rho_{++}(t)+\gamma_{\uparrow}\rho_{--}(t)+\bar{\gamma}\rho_{-+}(t)+\bar{\gamma}^*\rho_{+-}(t),\\
			\partial_t\rho_{+-}(t)&=-[\gamma_d+i\bar{\eta}]\rho_{+-}(t)+k_1\rho_{-+}(t)+k_-\rho_{--}(t)+k_+^*\rho_{++}(t),
		\end{align}
	\end{subequations}
	and $\partial_t\rho_{--}(t)=-\partial_t\rho_{++}(t)$ and $\partial_t\rho_{-+}(t)=\partial_t\rho_{-+}(t)^\dagger$. As defined in the main text, the secular rates are 
	\begin{align}
		\gamma_{\substack{\downarrow\\\uparrow}}&=\gamma_{\mp\mp}(\pm \eta),\\
		\gamma_d&=\tfrac{1}{2}\left[\gamma_{\uparrow}+\gamma_{\downarrow}\right]+2\gamma_{zz}(0),\\
		\bar{\eta}&=\eta+\Im\left[\Gamma_{--}(\eta)-\Gamma_{++}(-\eta)\right].
	\end{align}
	The rates are written in terms of the Markovian ECFs as
	\begin{align}
		\Gamma_{\alpha\beta}(\omega)=\intf{0}{\infty}{s}\rme^{i\omega s}\left\langle g_{\alpha}^\dagger(s) g_{\beta}(0)\right\rangle,
	\end{align}
	for $\alpha\in\{z,+,-\}$, and it is the expressions for $g_\alpha$ that vary between the PMFE and DFME, and these are given in the main text in Eqs.~\eqref{eq:g} and Appendix~\ref{app:DFME}, respectively. The non-secular rates are
	\begin{subequations}
		\begin{align}
			\bar{\gamma}&=\Gamma_{-z}(0)+\Gamma_{+z}(0)^*\\
			k_1&=\Gamma_{-+}(-\eta)+\Gamma_{+-}(\eta)^*,\\
			k_\pm&=\mp\Gamma_{\pm z}(0)\pm\Gamma_{\mp z}(0)^*\pm 2\Gamma_{z\mp}(\pm \eta)
		\end{align}
	\end{subequations}
	We evaluate $\Gamma_{\alpha\beta}(\omega)$ for the polaron frame in Appendix~\ref{app:gagb} and for the displaced frame in Appendix~\ref{app:DFME}.
	
	Notice that if the coherences are initially zero ($\rho_{+-}(0)=\rho_{-+}(0)=0$) then if $k_-=k_+=0$ as well, the coherences will be zero at all times $t$. This occurs in the polaron frame if $V\to0$ since $g_z\propto \sin(\varphi)\to 0$, i.e. the eigenstates fully localise, but this does not happen in the displaced frame in the same limit because $g_z^d$ does not become zero.

	\clearpage

	\section{Calculation of the polaron frame two-time correlation functions}\label{app:gagb}
	The environment correlation functions, which determine the second order Born-Markov rates, depend on the two-time correlation functions $\langle g^\dagger_\alpha(s)g_\beta(0)\rangle$ for $\alpha,\beta\in\{z,+,-\}$ where recalling from Eq.~\eqref{eq:g}, 
	\begin{subequations}
		\begin{align}
			g_z=&\ \frac{1}{2}\sin\left(\varphi\right)\left[\mathcal{C}+\mathcal{C}^\dagger\right],\\
			g_+=&\ \left[\cos^2\left(\frac{\varphi}{2}\right) \mathcal{C}-\sin^2\left(\frac{\varphi}{2}\right) \mathcal{C}^\dagger \right],
		\end{align}
	\end{subequations}
	and $g_-=g_+^\dagger$ where $\mathcal{C}=C-\kappa |V|$ and 
	\begin{equation}
		C=B_+\left(\sum_k\left[\mu\left(a^\dagger_k+\frac{\Delta^*_k}{\nu\uk}\right)+\bar{\mu}_k^*\left(a_k+\frac{\Delta_k}{\nu\uk}\right)\right]+|V|\right),
	\end{equation}
	and $B_\pm\equiv B(\pm2\delta)$ with $\delta_k=\Delta_k/\nu\uk$. The interaction picture form of the relevant operators are,
	\begin{align}
		a_k(s)&=a_k\rme^{-i\nu\uk s},\\
		B_\pm(s)&=B\left(\pm 2\delta\rme^{i\nu s}\right).
	\end{align}
	
	\subsection{Preliminary calculations}
	All $\langle g^\dagger_\alpha(s)g_\beta(0)\rangle$ depend on four two-time correlation functions: $\langle C^\dagger(s)C(0)\rangle$, $\langle C(s)C^\dagger(0)\rangle$, $\langle C(s)C(0)\rangle$, and $\langle C^\dagger(s)C^\dagger(0)\rangle$. Each of these in turn depends on either six or nine unique two-time correlation functions involving $a_k$, $a_k^\dagger$ and $B_\pm$. In this appendix, we list the results of all necessary two-time correlation functions which are each derived using the same mathematics as in the explicit examples in Appendix~\ref{app:Cexp}.
	
	\subsubsection{$\left\langle \mathcal{C}^\dagger(s)\mathcal{C}(0)\right\rangle$}
	To calculate this two-time correlation function, we require:
	\begin{subequations}\label{eq:CdCfuncs}
		\begin{align}
			\left\langle B_-(s)B_+(0)\right\rangle&=\kappa^2\rme^{\phi},\\
			\left\langle a_k^\dagger B_-(s)B_+(0)\right\rangle&=-N\uk x_k^*\kappa^2\rme^{\phi},\label{eq:CdCfuncsb}\\
			\left\langle a _kB_-(s)B_+(0)\right\rangle&=\tilde{N}\uk x_k\kappa^2\rme^{\phi},\\	
			\left\langle B_-(s)B_+(0)a_k^\dagger\right\rangle&=-\tilde{N}\uk x_k^*\kappa^2\rme^{\phi},\\	
			\left\langle B_-(s)B_+(0)a_k\right\rangle&=N\uk x_k\kappa^2\rme^{\phi},\\	
			\left\langle a_k^\dagger B_-(s)B_+(0)a_q^\dagger\right\rangle&=N\uk\tilde{N}\uq x_k^{*}x_q^*\kappa^2\rme^{\phi},\\	
			\left\langle a_k B_-(s)B_+(0)a_{q}\right\rangle&=\tilde{N}\uk N\uq x_kx_q\kappa^2\rme^{\phi},\\	
			\left\langle a_k^\dagger B_-(s)B_+(0)a_q\right\rangle&=\left(-N\uk N\uq x_k^*x_q+N\uk\delta_{kq}\right)\kappa^2\rme^{\phi},\label{eq:mm8}\\	
			\left\langle a_k B_-(s)B_+(0)a_q^\dagger\right\rangle&=\left(-\tilde{N}\uk \tilde{N}\uq x_kx_q^*+\tilde{N}\uk\delta_{kq}\right)\kappa^2\rme^{\phi},\label{eq:mm9}
		\end{align}
	\end{subequations}
	where, for brevity, $N\uk\equiv N(\nu\uk)$ is the Bose-Einstein distribution, $\tilde{N}\uk\equiv \tilde{N}(\nu\uk)=1+N(\nu\uk)$, $\phi\equiv\phi(s)$ given in Eq.~\eqref{eq:phi}, $\kappa^2=\exp[-\phi(0)]$, and finally
	\begin{equation}
		x_k\equiv x_k(s)=2\delta_k\left(1-\rme^{i\nu\uk s}\right).
	\end{equation}
	Substituting Eqs.~\eqref{eq:CdCfuncs} into the two-time correlation function we find that,
	\begin{align}\label{eq:CdC1}
		\left\langle \mathcal{C}^\dagger(s)\mathcal{C}(0)\right\rangle=\kappa^2\rme^{\phi(s)}\bigg[&\sum_k\left(\left|\bar{\mu}_k\right|^2N\uk\rme^{i\nu\uk s}+\left|\mu\uk\right|^2\tilde{N}\uk\rme^{-i\nu\uk s}\right)+\sum_{n,m\in\{-1,1\}}\sum_{kq}\chi_{nm}(k,q)\rme^{i n\nu\uk s}\rme^{i m\nu\uq s}\nonumber\\
		&+\sum_{n\in\{-1,1\}}\sum_k \xi_n(k)\rme^{in \nu\uk s}\bigg]+\kappa^2|V|^2\left(\rme^{\phi(s)}-1\right).
	\end{align}
	$\chi_{nm}(k,q)$ and $\xi_{n}(k)$ are written in matrix notation for brevity, using the format
	\begin{align}
		\chi(k,q)&=\begin{pmatrix}
			\chi_{-1,-1}(k,q) & \chi_{-1,1}(k,q)\\
			\chi_{1,-1}(k,q) & \chi_{1,1}(k,q)
		\end{pmatrix},\\
		\xi(k)&=\{\xi_{-1}(k),\xi_{1}(k)\}.
	\end{align}
	Written in this notation, one finds that
	\begin{subequations}\label{eq:AkqD}
		\begin{align}
			\chi(k,q)&=4\frac{f\uk^2}{\nu\uk}\frac{f\uq^2}{\nu\uq}\mu_{k}^{0*}\mu_{q}^0\Delta_{k}^0\Delta_{q}^0
			\begin{pmatrix}
				\tilde{N}\uk \tilde{N}\uq &  -\tilde{N}\uk  N_q \\
				-N\uk\tilde{N}\uq  & N\uk N\uq
			\end{pmatrix},\\
			\xi(k)&=2 |V| \frac{f_k^2}{\nu\uk}\Delta_k^0\left(\mu_k^0+\mu_k^{0*}\right)\{\tilde{N}\uk,-N\uk\},
		\end{align}
	\end{subequations}
	where we have written $\mu_k=i\mu_{k}^0 f\uk$, $\bar{\mu}_k=i\mu_{k}^{0*}f\uk$,  and= $\Delta_k=i\Delta_{k}^0f\uk$ with $\mu_k^0=\mb{d}_\mu\cdot\mb{e}_k$ and $\Delta_k^0=\mb{d}_\Delta\cdot\mb{e}_k$. 
	
	If one ignores the factor of $\kappa^2\exp[\phi(s)]$, the first term in Eq.~\eqref{eq:CdC1} will lead to the standard optical master equation rates describing photon emission and absorption scaling with the magnitude squared of the transition dipole strength. The second term in Eq.~\eqref{eq:CdC1} describes corrections to the standard optical master equation rates that are second order in both the permanent dipole moments and the transition dipole moments. The factor of $\kappa^2\exp[\phi(s)]$ accounts for higher order permanent dipole contributions. The second and third terms in Eq.~\eqref{eq:CdC1} along with Eqs.~\eqref{eq:AkqD} reveals the physical processes induced by the permanent dipoles. In the term dependent on $\chi_{nm}(k,q)$, each term $n$ and $m$ describes two simultaneous processes into the $k$ and $q$ modes, respectively. When $n$ or $m$ equal $-1$ the processes are photon emission, when $n$ or $m$ equal $0$ the processes are non-radiative, and when $n$ or $m$ equal $+1$ the processes are photon absorption. This can be seen by both the frequency dependence of the exponential phase factors in Eq.~\eqref{eq:CdC1} and by the dependence of the matrix elements on $\tilde{N}\uk$, $N\uk$, or neither. Similarly, the term dependent on $\xi_n(k)$ describes new one photon emission ($n=-1$) and absorption ($n=+1$) processes induced by the combination of the driving, permanent dipoles and transition dipoles.
	
	Finally, we take the continuum limit of the photon wavenumber to obtain
	\begin{align}\label{eq:CdC2}
		\left\langle \mathcal{C}^\dagger(s)\mathcal{C}(0)\right\rangle=&\kappa^2\rme^{\phi(s)}\bigg[\Omega_{\mu\bar{\mu}}\intf{0}{\infty}{\nu}J(\nu)\left(N(\nu)\rme^{i\nu s}+\tilde{N}(\nu)\rme^{-i\nu s}\right)+\cos^2\left(\theta_{\mu\Delta}\right)\sum_{n,m\in\{-1,1\}}\tilde{\chi}^{\mu\bar{\mu}}_{nm}(s)\nonumber\\
		&+2|V|\cos(\theta_{\mu\Delta})\left(\Omega_{\mu\Delta}+\Omega_{\bar{\mu}\Delta}\right)\intf{0}{\infty}{\nu}\frac{J(\nu)}{\nu}\left(-N(\nu)\rme^{i\nu s}+\tilde{N}(\nu)\rme^{-i\nu s}\right)\bigg]+\kappa^2 |V|^2\left(\rme^{\phi(s)}-1\right),
	\end{align}
	where we have defined the two-dimensional Fourier transform
	\begin{equation}\label{eq:2DFT}
		\tilde{\chi}^{ab}_{nm}(s)=\intf{0}{\infty}{\nu}\intf{0}{\infty}{\nu'}\chi^{ab}_{nm}(\nu,\nu')\rme^{i n \nu s}\rme^{im\nu' s}.
	\end{equation}
	The continuum limit of the coefficients can be written in the matrix representation as
	\begin{equation}\label{eq:Anunup1}
		\chi^{ab}(\nu,\nu')=4\frac{J(\nu)}{\nu}\frac{J(\nu')}{\nu'}\Omega_{a\Delta}\Omega_{b\Delta}\begin{pmatrix}
			\tilde{N}(\nu) \tilde{N}(\nu')  & -\tilde{N}(\nu) N(\nu') \\
			-N(\nu)\tilde{N}(\nu') & N(\nu) N(\nu')
		\end{pmatrix},
	\end{equation}
	where $a,b\in\{\mu,\bar{\mu}\}$. We have introduced the $a,b$ superscripts so that we may write all four two-time correlation functions of $ \mathcal{C}(s)$ in a unified notation. Note that $\chi^{\mu\bar{\mu}}(\nu,\nu')$ is the same as Eq.~\eqref{eq:Adagdot} in the main text. The Fourier transform of the first term ($\propto\Omega_{\mu\bar{\mu}}$) in Eq.~\eqref{eq:CdC2} leads to $\Gamma_1(\omega)$ in the main text, of the second term ($\propto\tilde{\chi}_{nm}^{\mu\bar{\mu}}(\nu,\nu')$) leads to $\Gamma_2(\omega)$, of the third term ($\propto |V|$) leads to $\Gamma_{V,1}(\omega)$, and of the final term ($\propto |V|^2$) leads to $\Gamma_{V,0}(\omega)$. 
	
	We now also introduce a more complicated notation for the Fourier transforms of the two-time correlation functions so that we can write all four down in a unified notation. Note that this notation differs again from the simpler notation in the main text. In the new notation, the Fourier transform of $\langle \mathcal{C}^\dagger(s)\mathcal{C}(0)\rangle$ is
	\begin{equation}\label{eq:GDc}
		\Gamma^{(\dagger,\cdot)}(\omega)\equiv\intf{0}{\infty}{\nu}\rme^{i\omega s}\left\langle \mathcal{C}^\dagger (s)\mathcal{C}(0)\right\rangle=\Gamma_1^{\mu\bar{\mu},+}(\omega)+\Gamma_2^{\mu\bar{\mu},+}(\omega)+\Gamma_{V,1}^{\mu\bar{\mu},+}(\omega)+\Gamma_{V,0}^{+}(\omega),
	\end{equation}
	where we have defined 
	\begin{subequations}\label{eq:apprates}
		\begin{align}
			\Gamma^{ab,\pm}_1(\omega)&=\pi\Omega_{ab}\intf{0}{\infty}{\nu} J(\nu)\left[N(\nu)\mathcal{K}^\pm(\omega+\nu)+\tilde{N}(\nu)\mathcal{K}^\pm(\omega-\nu)\right],\\
			\Gamma^{ab,\pm}_2(\omega)&=\kappa^2\cos^2\left(\theta_{\mu\Delta}\right)\sum_{n,m\in\{-1,1\}}\intf{0}{\infty}{s}\rme^{i\omega s}\rme^{\pm \phi(s)}\tilde{\chi}_{n,m}^{ab}(\nu,\nu'),\\
			\Gamma^{ab,\pm}_{V,1}(\omega)&=4\pi\left(\Omega_{a\Delta}+\Omega_{b\Delta}\right)|V|\cos\left(\theta_{\mu\Delta}\right)\intf{0}{\infty}{\nu}\frac{J(\nu)}{\nu}\left[\tilde{N}(\nu)\mathcal{K}^\pm(\omega-\nu)-N(\nu)\mathcal{K}^\pm(\omega+\nu)\right],\\
			\Gamma^{\pm}_{V,2}&=\kappa^2|V|^2 \left(\rme^{\pm\phi(s)}-1\right),
		\end{align}
	\end{subequations}
	and
	\begin{equation}
		\mathcal{K}^\pm(\varepsilon)=\frac{1}{\pi}\intf{0}{\infty}{s}\rme^{i\varepsilon s}\rme^{\pm\phi(s)-\phi(0)},
	\end{equation}
	is a generalisation of $\mathcal{K}(\varepsilon)$ in Eq.~\eqref{eq:K}. Note that the single mode truncation solution of $\mathcal{K}^-(\varepsilon)$ is obtained from the solution for $\mathcal{K}^+(\varepsilon)$ in the main text by replacing $W_n=S_s^n\exp[-S_s]/n!$ in Eq.~\eqref{eq:Al} with $W_n'=(-S_s)^n\exp[-S_s]/n!$ \cite{rouse2022analytic}.

	\subsubsection{$\left\langle \mathcal{C}(s)\mathcal{C}^\dagger(0)\right\rangle$}
	To calculate this two-time correlation function, we require:
	\begin{subequations}\label{eq:CCdfuncs}
		\begin{align}
			\left\langle B_+(s)B_-(0)\right\rangle&=\kappa^2\rme^{\phi},\\
			\left\langle B_+(s)a_k^\dagger B_-(0)\right\rangle&=\left(-2\delta_k^*-\tilde{N}\uk y_k^*\right)\kappa^2\rme^{\phi},\\
			\left\langle B_+(s)a_k B_-(0)\right\rangle&=\left(-2\delta_k+N\uk y_k\right)\kappa^2\rme^{\phi},\\
			\left\langle B_+(s)a_k^\dagger a_q^\dagger B_-(0)\right\rangle&=\left(-2\delta_k^*-\tilde{N}\uk y_k^*\right)\left(-2\delta_q^*-\tilde{N}\uq y_q^*\right)\kappa^2\rme^{\phi},\\
			\left\langle B_+(s)a_k a_q B_-(0)\right\rangle&=\left(-2\delta_k+N\uk y_k\right)\left(-2\delta_q+N\uq y_q\right)\kappa^2\rme^{\phi},\\
			\left\langle B_+(s)a_k^\dagger a_q B_-(0)\right\rangle&=\left[N\uk\delta_{kq}+\left(-2\delta_k^*-\tilde{N}\uk y_k^*\right)\left(-2\delta_q+N\uq y_q\right)\right]\kappa^2\rme^{\phi},\\
			\left\langle B_+(s)a_k a_q^\dagger B_-(0)\right\rangle&=\left[\tilde{N}\uk\delta_{kq}+\left(-2\delta_k+N\uk y_k\right)\left(-2\delta_q^*-\tilde{N}\uq y_q^*\right)\right]\kappa^2\rme^{\phi},
		\end{align}
	\end{subequations}
	where
	\begin{equation}
		y_k\equiv y_k(s)=2\delta_k\left(\rme^{i\nu\uk s}-1\right).
	\end{equation}
	
	Substituting Eqs.~\eqref{eq:CCdfuncs} into the two-time correlation function and taking the continuum limit one obtains
	\begin{align}\label{eq:CCd2}
		\left\langle \mathcal{C}(s)\mathcal{C}^\dagger(0)\right\rangle=\kappa^2&\rme^{\phi(s)}\bigg[\Omega_{\mu\bar{\mu}}\intf{0}{\infty}{\nu}J(\nu)\left(N(\nu)\rme^{i\nu s}+\tilde{N}(\nu)\rme^{-i\nu s}\right)+\cos^2(\theta_{\mu\Delta})\sum_{n,m\in\{-1,1\}}\tilde{\chi}_{nm}(s)\nonumber\\
		&-2|V|\cos(\theta_{\mu\Delta})\left(\Omega_{\mu\Delta}+\Omega_{\bar{\mu}\Delta}\right)\intf{0}{\infty}{\nu}\frac{J(\nu)}{\nu}\left(-N(\nu)\rme^{i\nu s}+\tilde{N}(\nu)\rme^{i\nu s}\right)\bigg]+\kappa^2 |V|^2\left(\rme^{\phi(s)}-1\right),
	\end{align}
	which is the same as Eq.~\eqref{eq:CdC2} except for the overall minus sign on the third term ($\propto V$). Written in terms of the rate functions defined in Eqs.~\eqref{eq:apprates} the Fourier transform of this two-time correlation function is,
	\begin{equation}\label{eq:GcD}
		\Gamma^{(\cdot,\dagger)}(\omega)\equiv\intf{0}{\infty}{\nu}\rme^{i\omega s}\left\langle \mathcal{C} (s)\mathcal{C}^\dagger(0)\right\rangle=\Gamma_1^{\mu\bar{\mu},+}(\omega)+\Gamma_2^{\mu\bar{\mu},+}(\omega)-\Gamma_{V,1}^{\mu\bar{\mu},+}(\omega)+\Gamma_{V,0}^{+}(\omega).
	\end{equation}

	\subsubsection{$\left\langle \mathcal{C}^\dagger(s)\mathcal{C}^\dagger(0)\right\rangle$}
	To calculate this two-time correlation function, we require:
	\begin{subequations}\label{eq:CdCdfuncs}
		\begin{align}
			\left\langle B_-(s)B_-(0)\right\rangle&=\kappa^2\rme^{-\phi},\\
			\left\langle a_k^\dagger B_-(s)B_-(0)\right\rangle&=-N\uk z_k^*\kappa^2\rme^{-\phi},\\
			\left\langle a_k B_-(s)B_-(0)\right\rangle&=\tilde{N}\uk z_k\kappa^2\rme^{-\phi},\\
			\left\langle B_-(s)a_k^\dagger  B_-(0)\right\rangle&=\left(-\tilde{N}\uk z_k^*-2\delta_k^*\right)\kappa^2\rme^{-\phi},\\
			\left\langle B_-(s)a_k B_-(0)\right\rangle&=\left(N\uk z_k-2\delta_k\right)\kappa^2\rme^{-\phi},\\
			\left\langle a_k^\dagger B_-(s)a_q^\dagger B_-(0)\right\rangle&=-N\uk z_k^*\left(-\tilde{N}\uq z_q^*-2\delta_q^*\right)\kappa^2\rme^{-\phi},\\
			\left\langle a_k B_-(s)a_q B_-(0)\right\rangle&=\tilde{N}\uk z_k\left(N\uq z_q-2\delta_q\right)\kappa^2\rme^{-\phi},\\
			\left\langle a_k^\dagger B_-(s)a_q B_-(0)\right\rangle&=\left[N\uk\delta_{kq}-N\uk z_k\left(N\uq z_q-2\delta_q\right)\right]\kappa^2\rme^{-\phi},\\
			\left\langle a_k B_-(s)a_q^\dagger B_-(0)\right\rangle&=\left[\tilde{N}\uk\delta_{kq}+\tilde{N}\uk z_k\left(N\uq z_q-2\delta_q\right)\right]\kappa^2\rme^{-\phi},
		\end{align}
	\end{subequations}
	where 
	\begin{equation}\label{eq:z}
		z_k\equiv z_k(s)=-2\delta_k\left(\rme^{i\nu\uk s}+1\right).
	\end{equation}
	Substituting Eqs.~\eqref{eq:CdCdfuncs} into the two-time correlation function and taking the continuum limit one obtains
	\begin{align}\label{eq:CdCd2}
		\left\langle \mathcal{C}^\dagger(s)\mathcal{C}^\dagger(0)\right\rangle=\kappa^2\rme^{-\phi(s)}\left[\Omega_{\bar{\mu}\bar{\mu}}\intf{0}{\infty}{\nu}J(\nu)\left(N(\nu)\rme^{i\nu s}+\tilde{N}(\nu)\rme^{-i\nu s}\right)-\sum_{n,m\in\{-1,1\}}\tilde{\chi}^{\bar{\mu}\bar{\mu}}_{nm}(s)\right]+\kappa^2|V|^2\left(\rme^{-\phi(s)}-1\right),
	\end{align}
	where $\tilde{\chi}_{nm}^{ab}(s)$ is defined in Eqs.~\eqref{eq:2DFT} and~\eqref{eq:Anunup1}. Eq.~\eqref{eq:CdCd2} has only a few differences to the two-time correlation functions $\langle \mathcal{C}^\dagger(s)\mathcal{C}(0)\rangle$ and $\langle \mathcal{C}(s)\mathcal{C}^\dagger(0)\rangle$ derived earlier. Most significantly, Eq.~\eqref{eq:CdCd2} has $\phi(s)\to-\phi(s)$, it does not have a term proportional to $|V|\cos(\theta_{\mu\Delta})$, and there is a minus sign on the second term. The final difference is that Eq.~\eqref{eq:CdCd2} only depends on $\mb{d}_{\bar{\mu}}=\mb{d}_\mu^*$.
	
	Using the functions defined in Eqs.~\eqref{eq:apprates} we can write the Fourier transform of Eq.~\eqref{eq:CdCd2} as
	\begin{equation}\label{eq:GDD}
		\Gamma^{(\dagger,\dagger)}(\omega)\equiv\intf{0}{\infty}{\nu}\rme^{i\omega s}\left\langle \mathcal{C}^\dagger (s)\mathcal{C}^\dagger(0)\right\rangle=\Gamma_1^{\bar{\mu}\bar{\mu},-}(\omega)-\Gamma_2^{\bar{\mu}\bar{\mu},-}(\omega)+\Gamma_{V,0}^{-}(\omega).
	\end{equation}

	\subsubsection{$\left\langle \mathcal{C}(s)\mathcal{C}(0)\right\rangle$}
	To calculate this two-time correlation function we require:
	\begin{subequations}\label{eq:CCfuncs}
		\begin{align}
			\left\langle B_+(s)B_+(0)\right\rangle&=\kappa^2\rme^{-\phi},\\
			\left\langle B_+(s) a_k^\dagger B_+(0)\right\rangle&=\left(z_k^*\tilde{N}\uk+2\delta_k^*\right)\kappa^2\rme^{-\phi},\\
			\left\langle B_+(s) a_k B_+(0)\right\rangle&=\left(-z_k N\uk+2\delta_k\right)\kappa^2\rme^{-\phi},\\
			\left\langle B_+(s)B_+(0)a_k^\dagger\right\rangle&=\tilde{N}\uk z_k^*\kappa^2\rme^{-\phi},\\
			\left\langle B_+(s)B_+(0)a_k\right\rangle&=-N\uk z_k\kappa^2\rme^{-\phi},\\
			\left\langle B_+(s)a_k^\dagger B_+(0)a_q^\dagger\right\rangle&=\tilde{N}\uk z_k^*\left(\tilde{N}\uq z_q^*+2\delta_q^*\right)\kappa^2\rme^{-\phi},\\
			\left\langle B_+(s)a_k B_+(0)a_q\right\rangle&=N\uk z_k\left(N\uq z_q-2\delta_q\right)\kappa^2\rme^{-\phi},\\
			\left\langle B_+(s)a_k^\dagger B_+(0)a_q\right\rangle&=\left[N\uk\delta_{kq}-\left(z_k^*\tilde{N}\uk+2\delta_k^*\right)N\uq z_q\right]\kappa^2\rme^{-\phi},\\
			\left\langle B_+(s)a_kB_+(0)a_q^\dagger\right\rangle&=\left[\tilde{N}\uk\delta_{kq}+\left(-z_kN\uk+2\delta_k\right)\tilde{N}_qz_q^*\right]\kappa^2\rme^{-\phi},
		\end{align}
	\end{subequations}
	where $z_k\equiv z_k(s)$ is given in Eq.~\eqref{eq:z}. Substituting Eqs.~\eqref{eq:CCfuncs} into the two-time correlation function and taking the continuum limit one obtains
	\begin{align}\label{eq:CC2}
		\left\langle \mathcal{C}(s)\mathcal{C}(0)\right\rangle=\kappa^2\rme^{-\phi(s)}\left[\Omega_{\mu\mu}\intf{0}{\infty}{\nu}J(\nu)\left(N(\nu)\rme^{i\nu s}+\tilde{N}(\nu)\rme^{-i\nu s}\right)-\sum_{n,m\in\{-1,1\}}\tilde{\chi}^{\mu\mu}_{nm}(s)\right]+\kappa^2|V|^2\left(\rme^{-\phi(s)}-1\right),
	\end{align}
	where $\tilde{\chi}_{nm}^{ab}(s)$ is defined in Eqs.~\eqref{eq:2DFT} and~\eqref{eq:Anunup1}. Eq.~\eqref{eq:CC2} is only different from Eq.~\eqref{eq:CdCd2} in that $\mb{d}_{\bar{\mu}}$ has been replaced with $\mb{d}_\mu$.
	
	Using the functions defined in Eqs.~\eqref{eq:apprates} we can write the Fourier transform of Eq.~\eqref{eq:CC2} as
	\begin{equation}\label{eq:Gcc}
		\Gamma^{(\cdot,\cdot)}(\omega)\equiv\intf{0}{\infty}{\nu}\rme^{i\omega s}\left\langle \mathcal{C} (s)\mathcal{C}(0)\right\rangle=\Gamma_1^{\mu\mu,-}(\omega)-\Gamma_2^{\mu\mu,-}(\omega)+\Gamma_{V,0}^{-}(\omega).
	\end{equation}
	
	\subsection{Two-time correlation functions $\left\langle g_\alpha^\dagger(s)g_\beta(0)\right\rangle$}
	With the expression for $\langle \mathcal{C}^\dagger(s)\mathcal{C}(0)\rangle$, $\langle \mathcal{C}(s)\mathcal{C}^\dagger(0)\rangle$, $\langle \mathcal{C}^\dagger(s)\mathcal{C}^\dagger(0)\rangle$ and $\langle \mathcal{C}(s)\mathcal{C}(0)\rangle$ given in Eqs.~\eqref{eq:CdC2}, \eqref{eq:CCd2}, \eqref{eq:CdCd2} and \eqref{eq:CC2} we can now write down the two-time correlation functions $\langle g_\alpha^\dagger(s)g_\beta(0)\rangle$ for $\alpha,\beta\in\{z,+,-\}$. We will write these expressions in the continuum limit using the generic operators
	\begin{equation}\label{eq:ga}
		g_\alpha=a_\alpha \mathcal{C} + b_\alpha \mathcal{C}^\dagger,
	\end{equation}
	and the equivalent for $\alpha\to\beta$. One can recover the desired two-time correlation functions of $g_-$, $g_+$ and $g_z$ by using the coefficients written in Table~\ref{tab:gs}. One finds that the Fourier transform of the two-time correlation functions are
	\begin{align}\label{eq:gab}
		\Gamma_{\alpha\beta}(\omega)\equiv\intf{0}{\infty}{s}\rme^{i\omega s}\left\langle g_{\alpha}^\dagger(s) g_{\beta}(0)\right\rangle=a_\alpha a_\beta \Gamma^{(\dagger,\cdot)}(\omega)+b_\alpha b_\beta \Gamma^{(\cdot,\dagger)}(\omega)+a_\alpha b_\beta \Gamma^{(\dagger,\dagger)}(\omega)+b_\alpha a_\beta \Gamma^{(\cdot,\cdot)}(\omega),
	\end{align}
	where $\Gamma^{(\dagger,\cdot)}(\omega)$, $\Gamma^{(\cdot,\dagger)}(\omega)$, $\Gamma^{(\dagger,\dagger)}(\omega)$, and $\Gamma^{(\cdot,\cdot)}(\omega)$ are given in Eqs.~\eqref{eq:GDc}, \eqref{eq:GcD}, \eqref{eq:GDD}, and \eqref{eq:Gcc}, respectively.
	\begin{table}[h!]
		\centering
		\begin{tabular}{ c | c c c } 
			\hline\hline
			$\alpha$ & $-$ & $+$ & $z$  \\ 
			\hline 
			$a_\alpha$ & $-\sin^2\left(\frac{\varphi}{2}\right)$& $\cos^2\left(\frac{\varphi}{2}\right)$ & $\sin^2\left(\frac{\varphi}{2}\right)\cos^2\left(\frac{\varphi}{2}\right)$\\
			$b_\alpha$ & $\cos^2\left(\frac{\varphi}{2}\right)$ & $-\sin^2\left(\frac{\varphi}{2}\right)$ & $\sin^2\left(\frac{\varphi}{2}\right)\cos^2\left(\frac{\varphi}{2}\right)$\\
			\hline\hline
		\end{tabular}
		\caption{\textbf{The coefficients used in $g_\alpha$ in Eq.~\eqref{eq:ga} to obtain the polaron frame coupling operators $g_-$, $g_+$ and $g_z$. Recall that $\cos^2(\varphi/2)=(1+\epsilon/\eta)/2$ and $\sin^2(\varphi/2)=(1-\epsilon/\eta)/2$.}}
		\label{tab:gs}
	\end{table}

	\clearpage

	\section{Correlation functions in the displaced frame master equation}\label{app:DFME}

	If one does not make the polaron transformation of the Hamiltonian in Eq.~\eqref{eq:Hd} and instead moves straight to the eigenbasis, one finds the displaced frame Hamiltonian
	\begin{equation}\label{eq:Hw}
		H=\frac{\eta^d}{2}\tau_z^d+\sum_k\nu\uk a_k^\dagger a_k+\sum_{\mu\in\{z,+,-\}}g^d_\mu\tau_\mu^d,	
	\end{equation}
	where $\eta^d=\sqrt{\epsilon^{2}+4|V|^2}$ [there is no $\kappa$ renormalisation of $V$] and 
	\begin{subequations}\label{eq:gweak}
		\begin{align}
			g_z^d=&\ \left[\cos^2\left(\frac{\varphi^d}{2}\right)-\sin^2\left(\frac{\varphi^d}{2}\right)\right]\pi_{\Delta\Delta}+\cos\left(\frac{\varphi^d}{2}\right)\sin\left(\frac{\varphi^d}{2}\right)\left[\pi_{\mu\bar{\mu}}+\pi_{\bar{\mu}\mu}\right]\\
			g_+^d=&-2\cos\left(\frac{\varphi^d}{2}\right)\sin\left(\frac{\varphi^d}{2}\right) \pi_{\Delta\Delta}+\cos^2\left(\frac{\varphi^d}{2}\right)\pi_{\mu\bar{\mu}}-\sin^2\left(\frac{\varphi^d}{2}\right)\pi_{\bar{\mu}\mu},
		\end{align}
	\end{subequations}
	and $g_-^d=g_+^{d \dagger}$, where
	\begin{equation}\label{eq:Cpq}
		\pi_{pq}=\sum_k\left(p_k a_k^\dagger + q_k^* a_k\right).
	\end{equation}
	The Pauli matrices in the eigenbasis are $\tau^d_{\pm}=\kb{\pm^d}{\mp^d}$ and $\tau^d_z=\kb{+^d}{+^d}-\kb{-^d}{-^d}$ where 
	\begin{equation}
		\begin{pmatrix}
			\ket{e} \\ \ket{g}
		\end{pmatrix}
		=
		\begin{pmatrix}
			\cos\left(\frac{\varphi^d}{2}\right) & -\sin\left(\frac{\varphi^d}{2}\right)\\
			\sin\left(\frac{\varphi^d}{2}\right)& \cos\left(\frac{\varphi^d}{2}\right)
		\end{pmatrix}
		\begin{pmatrix}
			\ket{+^d} \\ \ket{-^d}
		\end{pmatrix},
	\end{equation}
	with $\cos(\varphi^d)=\epsilon/\eta^d$ and $\sin(\varphi^d)=2|V|/\eta^d$.
	
	Since Eq.~\eqref{eq:Hw} has the same structure as Eq.~\eqref{eq:Hp}, the master equation in the displaced frame has the same algebraic form as in the polaron frame, i.e. given within the secular approximation by Eqs.~\eqref{eq:nonsec} and in full in Appendix~\ref{app:NS}, but the environment correlation functions (ECFs) are different. The displaced frame ECFs depend on linear combinations of Fourier transforms of the form
	\begin{equation}
		\Gamma^d_{pq,rs}(\omega)=\intf{0}{\infty}{s}\rme^{i\omega s}\left\langle \pi_{pq}^\dagger(s)\pi_{rs}(0)\right\rangle\equiv\frac{1}{2}\gamma_{pq,rs}^d(\omega)+iS_{pq,rs}^d(\omega).
	\end{equation}
	Substituting Eq.~\eqref{eq:Cpq} into the ECF and using that $\langle a_k^\dagger a_{k'}\rangle = \delta_{kk'}N(\nu\uk)$,  $\langle a_k a_{k'}^\dagger\rangle = \delta_{kk'}\tilde{N}(\nu\uk)$ where $\tilde{N}(\nu)=1+N(\nu)$ and that other combinations equal zero, leads to
	\begin{subequations}\label{eq:weakrate}
		\begin{align}
			\gamma^d_{pq,rs}(\omega)&=2\pi\left[\cos\left(\theta_{pr}\right)\Omega_{\bar{p}r}J(\omega)\tilde{N}(\omega)+\cos\left(\theta_{qs}\right)\Omega_{q\bar{s}}J(-\omega)N(-\omega)\right],\label{eq:weakratea}\\
			S^d_{pq,rs}(\omega)&=\mathcal{P}\intf{0}{\infty}{\nu}J(\nu)\left[\cos\left(\theta_{pr}\right)\Omega_{\bar{p}r}\frac{\tilde{N}(\nu)}{\omega-\nu}+\cos\left(\theta_{qs}\right)\Omega_{q\bar{s}}\frac{N(\nu)}{\omega+\nu}\right].
		\end{align}
	\end{subequations}	
	
	To exemplify how these functions relate to the rates in Eqs.~\eqref{eq:nonsec} we will derive $\gamma_{--}^d(\eta^d)$ explicitly. This function is given by
	\begin{equation}
		\gamma_{--}^d(\eta^d)=2\Re\intf{0}{\infty}{s}\rme^{i\eta^ds}\left\langle g_-^{d\dagger}(s)g_-^d(0)\right\rangle.
	\end{equation}
	Substituting in $g^d_-$ and using Eqs.~\eqref{eq:weakrate} we can read off that [ignoring arguments $(\eta^d)$ on the right-hand-side and temporarily denoting $\cos(\varphi^d/2)=c$ and $\sin(\varphi^d/2)=s$],
	\begin{align}
		\gamma_{--}^d(\eta^d)=&\ 4c^2s^2\gamma^d_{\Delta\Delta,\Delta\Delta}+2cs^3\rme^{i\theta}\gamma^d_{\Delta\Delta,\mu\bar{\mu}}-2c^3s\rme^{-i\theta}\gamma^d_{\Delta\Delta,\bar{\mu}\mu}\nonumber\\
		&+2cs^3\rme^{-i\theta}\gamma^d_{\mu\bar{\mu},\Delta\Delta}+s^4\gamma^d_{\mu\bar{\mu},\mu\bar{\mu}}-s^2c^2\rme^{-2i\theta}\gamma^d_{\mu\bar{\mu},\bar{\mu}\mu}\nonumber\\
		&-2c^3s\rme^{i\theta}\gamma^d_{\bar{\mu}\mu,\Delta\Delta}-c^2s^2\rme^{2i\theta}\gamma_{\bar{\mu}\mu,\mu\bar{\mu}}^d+c^4\gamma^d_{\bar{\mu}\mu,\bar{\mu}\mu}.
	\end{align}
	Using the symmetry that $\Omega_{ab}=\Omega_{ba}$ and that $\mb{d}_\Delta\in\Re$, we can read off from Eq.~\eqref{eq:weakratea} that
	\begin{align}
		\gamma^d_{\Delta\Delta,\Delta\Delta}(\eta^d)&=\Omega_{\Delta\Delta}\gamma_0(\eta^d),\\
		\gamma^d_{\bar{\mu}\mu,\mu\bar{\mu}}(\eta^d)&=\Omega_{\mu\mu}\gamma_0(\eta^d),\\
		\gamma^d_{\mu\bar{\mu},\bar{\mu}\mu}(\eta^d)&=\Omega_{\bar{\mu}\bar{\mu}}\gamma_0(\eta^d),\\
		\gamma^d_{\bar{\mu}\mu,\Delta\Delta}(\eta^d)&=\gamma^d_{\Delta\Delta,\mu\bar{\mu}}(\eta^d)=\cos\left(\theta_{\mu\Delta}\right)\Omega_{\mu\Delta}\gamma_0(\eta^d),\\
		\gamma^d_{\mu\bar{\mu},\Delta\Delta}(\eta^d)&=\gamma^d_{\Delta\Delta,\bar{\mu}\mu}(\eta^d)=\cos\left(\theta_{\mu\Delta}\right)\Omega_{\bar{\mu}\Delta}\gamma_0(\eta^d),\\
		\gamma^d_{\mu\bar{\mu},\mu\bar{\mu}}(\eta^d)&=\gamma^d_{\bar{\mu}\mu,\bar{\mu}\mu}(\eta^d)=\cos\left(\theta_{\mu\Delta}\right)\Omega_{\mu\bar{\mu}}\gamma_0(\eta^d),
	\end{align}
	where $\gamma_0(\omega)=2\pi[J(\omega)\tilde{N}(\omega)+J(-\omega)N(-\omega)$]. Using these we find that
	\begin{equation}
		\gamma_{--}^d(\eta^d)=k^0_{--}\gamma_0(\eta^d),	
	\end{equation}
	where the rate coefficient is
	\begin{equation}\label{eq:Gmm}
		k^0_{--}=\left[c^4+s^4\right]\Omega_{\mu\bar{\mu}}+4c^2s^2\Omega_{\Delta\Delta}-s^2c^2\left[\Omega_{\bar{\mu}\bar{\mu}}+\Omega_{\mu\mu}\right]-2s c\left[c^2-s^2\right]\left[\Omega_{\bar{\mu}\Delta}+\Omega_{\mu\Delta}\right]\cos\left(\theta_{\mu\Delta}\right).
	\end{equation}
	This is rather complicated but only because we have complex transition dipole moments. If we assume real transition dipoles ($\mb{d}_{\bar{\mu}}\to \mb{d}_\mu$) we find $k^0_{--}=\Omega_{--}=(8\pi/3)(\mb{d}_-\cdot\mb{d}_-)$ where we have defined a new dipole vector
	\begin{equation}
		\mb{d}_-=\left[c^2-s^2\right]\Re[\mb{d}_\mu]+2cs\mb{d}_\Delta.
	\end{equation}
	Likewise, if we choose purely imaginary transition dipoles ($\mb{d}_\mu\to i\Im[\mb{d}_\mu]$) we find that $k^0_{--}=\Omega_{--}$ but with the dipole vector now defined as
	\begin{equation}
		\mb{d}_-=\left[c^2-s^2\right]\Im[\mb{d}_\mu]+2cs\mb{d}_\Delta.
	\end{equation}
	A final limit worth checking is when the eigenstates fully localise, $\ket{+}\to\ket{e}$ and $\ket{-}\to\ket{g}$, i.e. $c\to1$ and $s\to 0$. In this case, $k_{--}^0\to\Omega_{\mu\bar{\mu}}$, and so the decay rate is $\gamma_{--}^d(\eta^d)=\gamma_{--}^d(\epsilon)=\Omega_{\mu\bar{\mu}}\gamma_0(\epsilon)$, which is the decay rate in the standard optical master equation.

	\clearpage

	\clearpage

	\section{Derivation of the emission spectrum}\label{app:spectrum}
	In this appendix, we derive Eq.~\eqref{eq:Semm}, which gives the emission spectrum for the single emitter system. Our derivation closely follows those provided for single emitter systems without permanent dipoles in Refs.~\cite{ficek2005quantum,roy2015quantum,iles2017phonon,hughes2009theory,walls1994gj}. Due to the permanent dipoles, there are additional light-matter interaction terms in the Hamiltonian in Eq.~\eqref{eq:Hl}, and so the derivation is slightly more cumbersome. However, as we show here, these additional terms do not affect the expression for the emission spectrum of the emitter within the standard approximations.
	
	The emission spectrum is given exactly by
	\begin{equation}\label{eq:Semm0}
		I(\omega)=\lim_{t\to\infty}\Re\intf{0}{\infty}{\tau}\rme^{i\omega\tau}\left\langle \mb{E}_-\left(\mb{R},t\right)\cdot\mb{E}_+\left(\mb{R},t+\tau\right)\right\rangle,
	\end{equation}
	where $\mb{R}$ is the position of the detector and the positive frequency component of the electric field is
	\begin{equation}\label{eq:Ep}
		\mb{E}_+\left(\mb{R},t\right)=i\sum_k\mb{e}_kf\uk a_k\rme^{-i\mb{k}\cdot\mb{R}},
	\end{equation}
	and $\mb{E}_-(\mb{R},t)=\mb{E}_+(\mb{R},t)^\dagger$. Our aim is to express the expectation value $\langle a_k^\dagger a_{q}\rangle$ in terms of dipole operators $\sigma_\pm$, which is achieved through the Heisenberg equation of motion
	\begin{equation}
		\frac{\partial}{\partial t}a_k(t)=-i\left[H,a_k(t)\right],
	\end{equation}
	where $H$ is the lab frame Hamiltonian in Eq.~\eqref{eq:H} and equivalently in Eq.~\eqref{eq:Hl}. In doing so we will have arrived at the form of the emission spectrum in Eq.~\eqref{eq:Semm}. 
	
	Using $[a_k,a_q^\dagger]=\delta_{kq}$, one can show that
	\begin{equation}\label{eq:dak}
		\frac{\partial}{\partial t}a_k(t)=-i\nu\uk a_k(t)-\Delta_k\sigma_z(t)-\mu_k\sigma_+(t)-\mu_k^*\sigma_-(t),
	\end{equation}
	where $\sigma_\alpha(t)=U(t)^\dagger\sigma_\alpha U(t)$ and $U(t)=\exp(-i Ht)$, which can be solved to yield
	\begin{equation}\label{eq:ak}
		a_k(t)=a_k(0)\rme^{-i\nu\uk t}-\intf{0}{t}{\tau}\rme^{i\nu\uk\left(\tau-t\right)}\left[\Delta_k\sigma_z(\tau)+\mu_k\sigma_+(\tau)-\mu_k^*\sigma_-(\tau)\right].
	\end{equation}
	The first term in Eq.~\eqref{eq:ak} is the free field term which does not contribute to the spectrum of the emitter and is henceforth ignored. After substituting the second term of Eq.~\eqref{eq:ak} into Eq.~\eqref{eq:Ep} and taking the continuum limit we obtain
	\begin{equation}\label{eq:Em2}
		\mb{E}_+\left(\mb{R},t\right)=\intf{0}{\infty}{\nu}\sqrt{J(\nu)}\intf{0}{t}{\tau}\rme^{i\nu\uk\left(\tau-t\right)}\left[\mb{O}_z\sigma_z(\tau)+\mb{O}_\mu\sigma_+(\tau)+\mb{O}_{\bar{\mu}}\sigma_-(\tau)\right],
	\end{equation}
	where
	\begin{equation}
		\mb{O}_p=\int_{d\Omega\uk}\ \text{d}\Omega\uk \sum_\lambda\mb{e}_k\left(\mb{d}_p\cdot\mb{e}_k\right)\rme^{i\mb{k}\cdot\left(\mb{r}-\mb{R}\right)}.
	\end{equation}

	To progress analytically, we need to know how the emitter operators $\sigma_\alpha(t)$ evolve in time. However, this is very complicated. Instead, we make the so-called harmonic decomposition (see Chapter 2.2 of Ref.~\cite{ficek2005quantum}) in which we assume that the timescale over which the emitter evolves unitarily is much faster than the timescale over which spontaneous emission occurs. Within this approximation, we write that
	\begin{equation}
		\sigma_+(\tau)\approx\sigma_+(t)\rme^{i\epsilon\left(\tau-t\right)},\quad\ \sigma_-(\tau)\approx\sigma_-(t)\rme^{-i\epsilon\left(\tau-t\right)},\quad \text{ and}\quad\ \sigma_z(\tau)\approx\sigma_z(t).
	\end{equation}
	Making the harmonic decomposition in Eq.~\eqref{eq:Em2} yields
	\begin{equation}\label{eq:Ep3}
		\mb{E}_+(\mb{R},t)=\intf{0}{\infty}{\nu}\sqrt{J(\nu)}\left[\mb{O}_z\sigma_z(t)j_0(\nu,t)+\mb{O}_\mu\sigma_+(t)j_+(\nu,t)+\mb{O}_{\bar{\mu}}\sigma_-(t)j_-(\nu,t)\right],
	\end{equation}
	where
	\begin{equation}
		j_\alpha(\nu,t)=\intf{0}{t}{\tau}\rme^{i\left(\tau-t\right)\left(\nu+\alpha\epsilon\right)}=-i\frac{1-\rme^{-it\left(\nu+\alpha\epsilon\right)}}{\nu+\alpha\epsilon}.
	\end{equation}

	The function $j_\alpha(\nu,t)$ is dominated by the contribution near to $\nu=-\alpha\epsilon$ and so we approximate it as a delta function, $j_\alpha(\nu,t)\approx2\pi\delta(\nu+\alpha\epsilon)$ \cite{ficek2005quantum}. After performing this approximation, one finds that the delta function corresponding to $j_+(\nu,t)$ lies outwith the integration domain $\nu\in[0,\infty]$ and so the term going as $\sigma_+(t)$ in Eq.~\eqref{eq:Ep3} vanishes. Moreover, the term proportional to $j_0(\nu,t)$ provides a delta function at zero frequency, leading to the term going as $\sigma_z(t)$ in Eq.~\eqref{eq:Ep3} to be proportional to $\sqrt{J(0)}$ which is equal to zero for any well-defined spectral density. Therefore, within the standard approximations outlined in this derivation, the presence of permanent dipoles does not change the expression of the emission spectrum from Eq.~\eqref{eq:Semm}. Thus, the only surviving term in Eq.~\eqref{eq:Ep3} is proportional to $\sigma_-(t)$, and so we obtain
	\begin{equation}\label{eq:Ep4}
		\mb{E}_+\left(\mb{R},t\right)\approx2\pi\sqrt{J(\epsilon)}\mb{O}_{\bar{\mu}}\left(\mb{r},\mb{R}\right)\sigma_-(t),
	\end{equation}
	where we have made the dependence of $\mb{O}_{\bar{\mu}}$ on $\mb{r}$ and $\mb{R}$ explicit. After substituting Eq.~\eqref{eq:Ep4} and its Hermitian conjugate into Eq.~\eqref{eq:Semm0}, one obtains Eq.~\eqref{eq:Semm}, and an explicit expression for the Green's function $\alpha_\text{prop}(\mb{r},\mb{R},
\omega)$.

	\clearpage
	
	\section{Effective Hamiltonian}
	\label{app:EffHam}
	In many numerical schemes to solve the open quantum dynamics of systems coupled to thermal environments, it is assumed that the environment is in a free Gibbs state at the associated temperature $T=1/\beta$. In the polaron framework, the environment is not in such a convenient form. As such, we derive here an effective Hamiltonian that encapsulates the polaron thermalised state but has the effective environment in a free Gibbs state. 
	
	We start with the lab frame Hamiltonian,
	\begin{equation}\label{eq:H2}
		H=\frac{\epsilon}{2}\sigma_z+V\sigma_++V^*\sigma_-+E_\text{dip}+\sum_k\nu\uk a_k^\dagger a_k+\pi_{DD}\mathcal{I}+\pi_{\Delta\Delta}\sigma_z+\pi_{\mu\bar{\mu}}\sigma_++\pi_{\bar{\mu}\mu}\sigma_-,
	\end{equation}
	where
	\begin{align}
		\pi_{pq}&=\sum_k\left(p_ka_k^\dagger+q_k^*a_k\right),\\
		p_k&=i\left(\mb{d}_p\cdot\mb{e}\uk\right),
	\end{align} 
	with $p,q\in\{\mu,\bar{\mu},D,\Delta\}$. In our calculations, the polaron frame initial state is 
	\begin{equation}
		\rho_p^0=\proj{g}{g}\otimes\rho_E,
	\end{equation}
	where $\rho_E=\exp[-\beta\sum_k\nu\uk a_k^\dagger a_k]/\mathcal{Z}_E$ is a thermal state. Performing the inversion of the unitary transformations to go from the displaced polaron frame to the displaced frame, the initial state is $\rho_d^0=U^\dagger \rho_p^0U $ where the polaron transformation is $U=B(\delta)\proj{e}{e}+B(-\delta)\proj{g}{g}$ and $\delta_k=\Delta_k/\nu\uk$ and $B(\alpha)=\exp[\sum_k(\alpha_k a_k^\dagger-\alpha_k^* a_k)]$. Therefore,
	\begin{equation}
		\rho_d^0=\proj{g}{g}\otimes\eta(\delta),
	\end{equation}
	where we have defined a displaced thermal state as $\eta(\alpha)=B(\alpha)\rho_E B(-\alpha)$. We can then obtain the lab frame initial state via $\rho^0_l=B(-d)\rho_d^0B(d)$ where $d_k=D_k/\nu\uk$, leading to
	\begin{equation}
		\rho^0_l=\proj{g}{g}\otimes \eta(\delta-d).
	\end{equation}
	The expression for $\eta(\alpha)$ can be rewritten by making use of the identity,
	\begin{equation}
		\exp\left(\rme^S X \rme^{-S}\right)=\rme^S\rme^X\rme^{-S},
	\end{equation}
	which holds if $\rme^S\rme^{-S}=\mathcal{I}$. Using this identity yields,
	\begin{equation}\label{eq:disp1}
		\eta(\alpha)=\exp\left[-\beta\sum_k\nu\uk\left(a_k^\dagger-\alpha_k^*\right)\left(a_k-\alpha_k\right)\right].	
	\end{equation}
	
	We now rewrite the lab frame Hamiltonian using new ladder operators $b_k$, which we will relate to the $a_k$ to ensure we model our desired initial state, $\eta(\delta-d)$. The initial environment state in the effective lab frame will be $\bar{\rho}_E=\exp[-\beta\sum_k\nu\uk b_k^\dagger b_k]/\mathcal{Z}_E$ and so by comparison with Eq.~\eqref{eq:disp1}, we know that
	\begin{equation}
		b_k=a_k-(\delta_k-d_k).
	\end{equation}
	Substituting this into our actual lab frame Hamiltonian in Eq.~\eqref{eq:H2} and ignoring any terms that are identity operators in both Hilbert spaces, we arrive at the effective Hamiltonian:
	\begin{equation}\label{eq:Hbar}
		\bar{H}=\bar{H}_0+\bar{H}_I,	
	\end{equation}
	where
	\begin{align}
		\bar{H}_0&=\left(\frac{\epsilon}{2}+G_{\Delta\Delta}\right)\sigma_z+V\sigma_++V^*\sigma_-+G_{\mu\bar{\mu}}\sigma_++G_{\bar{\mu}\mu}\sigma_-+\sum_k\nu\uk b_k^\dagger b_k,\\
		\bar{H}_I&=\pi'_{\Delta\Delta} \left(\mathcal{I}+\sigma^z\right)+\pi'_{\mu\bar{\mu}}\sigma_++\pi'_{\bar{\mu}\mu}\sigma_-,
		\label{eqn:EffHam}
	\end{align}
	and
	\begin{align}
		\pi'_{pq}&=\sum_k\left(p_k b_k^\dagger+q_k^* b_k\right),\\
		G_{pq}&=\sum_k\left(p_k\delta_k^*+q_k^*\delta_k\right).
	\end{align}
	In order to get numerical agreement between TEMPO and the PFME, both of which assume separable initial states but crucially in different frames, we, therefore, must use $\bar{H}$ in Eq.~\eqref{eq:Hbar} for TEMPO calculations.

\end{document}